\documentclass[a4paper,11pt]{article}
\pdfoutput=1
\usepackage{jcappub}
\usepackage[T1]{fontenc} 
\usepackage[usenames,dvipsnames,table]{xcolor}

\usepackage{multirow}
\usepackage{makecell}
\usepackage{pdflscape}
\usepackage{graphicx}
\usepackage{xspace}
\usepackage{tcolorbox}

\def\lsim{\mathrel{\raise.3ex\hbox{$<$\kern-.75em\lower1ex\hbox{$\sim$}}}}
\def\gsim{\mathrel{\raise.3ex\hbox{$>$\kern-.75em\lower1ex\hbox{$\sim$}}}}

\newcommand{\GeV}{\mathrm{GeV}}
\newcommand{\MeV}{\mathrm{MeV}}
\newcommand{\eV}{\mathrm{eV}}

\newcommand{\beq}{\begin{equation}}
\newcommand{\eeq}{\end{equation}}
\newcommand{\ba}{\begin{align}}
\newcommand{\ea}{\end{align}}

\newcommand{\Neff}{N_\mathrm{eff}}
\newcommand{\DNeff}{\Delta N_\mathrm{eff}}

\newcommand{\Mpc}{\mathrm{Mpc}}

\newcommand{\yr}{\mathrm{yr}}

\newcommand{\pd}[2]{\frac{\partial #1}{\partial #2}}

\newcommand{\lcdm}{\Lambda\mathrm{CDM}}

\newcommand{\CLASS}{\texttt{CLASS}\xspace}
\newcommand{\MP}{\texttt{MontePython}\xspace}

\newcommand{\dr}{\mathrm{dr}}

\newcommand{\rdr}{\rho_{\mathrm{dr}}}
\newcommand{\Hcon}{\mathcal{H}}

\newcommand{\Neffx}{N_{\mathrm{eff},\,x}}
\def\lcdm{$\Lambda$CDM\xspace}
\newcommand\KICP{Kavli Institute for Cosmological Physics, University of Chicago, Chicago, IL 60637, USA}
\newcommand\FNAL{Fermi National Accelerator Laboratory, Batavia, IL 60510, USA}
\newcommand\astro{Department of Astronomy and Astrophysics, University of Chicago, Chicago, IL 60637, USA}

\proceeding{\hfill FERMILAB-PUB-20-152-A}

\title{Warm Decaying Dark Matter and the Hubble Tension}

\author[a,b]{Nikita Blinov,}
\author[b,c]{Celeste Keith,}
\author[a,b,c]{Dan Hooper}
\affiliation[a]{\FNAL}
\affiliation[b]{\KICP}
\affiliation[c]{\astro}
\emailAdd{nblinov@fnal.gov}
\emailAdd{ckeith@uchicago.edu}
\emailAdd{dhooper@fnal.gov}
\abstract{

If a fraction of the dark matter is unstable and decays into dark radiation at around the time of matter-radiation equality, it could impact the expansion history of the universe in a way that helps to ameliorate the long-standing tension between the locally measured value of the Hubble constant and the value inferred from measurements of the cosmic microwave background and baryon acoustic oscillations (assuming standard $\Lambda$CDM cosmology). If this component of decaying dark matter is cold, however, it will modify the evolution of the gravitational potentials, leading to inconsistencies with these same data sets. With this in mind, we consider here a component of decaying warm dark matter, with a free-streaming length that is long enough to remain consistent with existing data. We study the background and perturbation evolution of warm decaying dark matter, and use cosmological data to constrain the mass, abundance and decay rate of such a particle. We find that a component of warm decaying dark matter can significantly reduce the tension between local and cosmological determinations of the Hubble constant.}

\begin{document}
\maketitle
\section{Introduction}

The Cosmic Microwave Background (CMB) provides us with
an exquisite probe of the universe prior to recombination. 
While the observed characteristics of the CMB are consistent with the predictions 
of the standard cosmological model, \lcdm, a number of tensions with local (or late-time) measurements have emerged in recent years. 
The most well-known of these tensions is the $\sim 4\sigma$ discrepancy 
between the value of the Hubble constant, $H_0$, as inferred from the CMB~\cite{Aghanim:2018eyx} or other independent early-time cosmological probes~\cite{Addison:2017fdm,Schoneberg:2019wmt,Philcox:2020vvt} (assuming standard \lcdm), and as directly measured in the local universe~\cite{Riess:2016jrr,Shanks:2018rka,Riess:2018kzi,Riess:2019cxk,Wong:2019kwg}. Note, however, that other local measurements of $H_0$~\cite{Freedman:2019jwv,Huang:2019yhh,Pesce:2020xfe} are compatible within uncertainties with \emph{both} the early time measurements and those of, e.g., Refs.~\cite{Riess:2019cxk,Wong:2019kwg}. In this work we assume that the true 
value of $H_0$ is larger than the one derived from the CMB as implied by the local observations, and seek a cosmological explanation for this discrepancy.
In addition to the $H_0$ tension, there is also a mild disagreement between the CMB-inferred~\cite{Aghanim:2018eyx} and directly measured values~\cite{Hikage:2018qbn,Abbott:2018wzc} 
of $\sigma_8$, the amplitude of matter fluctuations on 
$8h^{-1}$ Mpc scales.

Unless the Hubble tension is a consequence of some yet-to-be identified systematic error, this discrepancy suggests a departure from standard \lcdm which reduces the sound horizon during the era leading up to recombination~\cite{Vonlanthen:2010cd, Bernal:2016gxb, Verde:2016wmz, Evslin:2017qdn, Aylor:2018drw}. 
Many mechanisms have been proposed which fall within this class of ``early-time'' solutions~\cite{Weinberg:2013kea,Lesgourgues:2015wza,Shakya:2016oxf,DiValentino:2017oaw, Poulin:2018dzj,DEramo:2018vss, Poulin:2018cxd,Dessert:2018khu,Bringmann:2018jpr,Pandey:2019plg,Agrawal:2019lmo, Escudero:2019gzq,Hooper:2019gtx,Escudero:2019gvw,Alcaniz:2019kah}, nearly all of which involve the presence of additional energy at around the time of matter-radiation equality.\footnote{Models with additional energy near matter-radiation equality typically do not ease the $\sigma_8$ tension, since they require a larger value of $\Omega_m$ to keep the redshift at matter-radiation equality fixed, which increases the value of $\sigma_8$. This correlation is lessened in models 
with non-free-streaming radiation~\cite{Brust:2017nmv,Blinov:2020hmc}, larger neutrino masses~\cite{Kreisch:2019yzn}, or dark matter that experiences drag~\cite{Buen-Abad:2015ova,Buen-Abad:2017gxg}.} This additional energy could take the form of dark radiation, contributing to the value of $\Neff$, or could instead evolve non-trivially with redshift. The dynamics associated with this energy could also change the evolution of cosmological perturbations, leading to observable consequences and allowing us to place constraints on such scenarios. One simple possibility would be to introduce a component of cold dark matter (CDM) that decays into dark radiation prior to recombination. Although this could potentially impact the expansion history in a way that can help to alleviate the Hubble tension, the extra CDM would modify the evolution of the gravitational potentials in a way that is not consistent with the measured characteristics of the CMB~\cite{Poulin:2016nat}. As a result, decaying CDM (DCDM) cannot be invoked to address the Hubble tension.

In light of these considerations, it is well-motivated to study a modified version of this scenario. Refs.~\cite{Vattis:2019efj,Haridasu:2020xaa}, for example, considered late-universe DCDM where the daughter particles are semi-relativistic, weakening the constraints from Ref.~\cite{Poulin:2016nat} on DCDM decaying to dark radiation. 
In this work we instead consider the case in which the decaying matter has an appreciable 
free-streaming length, and therefore does not cluster on small scales as CDM does. 
A component of decaying \emph{warm} dark matter (DWDM) can be easily realized in simple and well-motivated particle physics models, such as a sterile neutrino with couplings to a dark sector. 
Such scenarios can arise within the context of neutrino mass generation~\cite{Gelmini:1980re,Chikashige:1980ui,Georgi:1981pg,Schechter:1981cv,Gelmini:1983ea,Gelmini:1984pe,Chacko:2019nej} and have been considered to explain accelerator anomalies~\cite{PalomaresRuiz:2005vf, Dentler:2019dhz}. 
From the point-of-view of cosmology, DWDM is a natural interpolation between two simple extensions of \lcdm: a completely relativistic dark sector 
modeled by $\DNeff$ (i.e., dark radiation), and decaying CDM.

Similar models have been studied before~\cite{Gelmini:1984pe,Turner:1984nf,1984MNRAS.211..277D,1988AZh....65..248D,Doroshkevich:1989bf,Bharadwaj:1997dz,Lopez:1998jt,Kaplinghat:1999xy,Lopez:1999ur,Chacko:2019nej}, especially 
in the context of decaying Standard Model (SM) neutrinos at late times.
In this paper we instead consider the dynamics of an additional component beyond the SM, such as a sterile neutrino, focusing on decays which occur prior to recombination. 
In Sec.~\ref{sec:model}, we discuss how such a scenario could arise in concrete particle physics models. 
We then derive the relevant background and perturbation equations, along with their initial 
conditions in Sec.~\ref{sec:boltzmann}. These are implemented in the Boltzmann solver \CLASS~\cite{Blas:2011rf,2011JCAP...09..032L}, which we use in combination with \MP~\cite{Audren:2012wb,Brinckmann:2018cvx} to perform a Monte Carlo study of the model parameter space 
in light of the latest results from Planck~\cite{Aghanim:2019ame}, baryon acoustic oscillation (BAO) data~\cite{Alam:2016hwk,Beutler:2011hx,Ross:2014qpa}, and Cepheid-calibrated 
distance ladder determinations of $H_0$~\cite{Riess:2019cxk}.\footnote{Other 
  local measurements of $H_0$, such as those described in Refs.~\cite{Wong:2019kwg,Huang:2019yhh,Freedman:2019jwv,Pesce:2020xfe}, are more consistent with the CMB-inferred value due to larger uncertainties or to a slightly lower central value than that of Ref.~\cite{Riess:2019cxk}. See Ref.~\cite{Verde:2019ivm} for a brief review. If these measurements are 
  used in place of Ref.~\cite{Riess:2019cxk} in the likelihood, the preference for non-standard cosmology would be weaker.} We find that the inclusion of a component of DWDM can significantly reduce the Hubble tension, from over 4$\sigma$ to approximately 2.9$\sigma$, after accounting for both CMB and BAO data. These results are 
described in Sec.~\ref{sec:results}, and our conclusions are summarized in Sec.~\ref{sec:conclusion}.

\section{Decaying Warm Dark Matter}
\label{sec:model}
The distinguishing feature of DWDM models is that 
the decaying fluid undergoes free-streaming which prevents 
clustering on length scales relevant to the CMB. We 
can determine what kinds of particles satisfy this condition 
by considering the characteristic free-streaming length~\cite{Kolb:1990vq}:
\beq
    \lambda_{\mathrm{fs}} = \int_0^{t_{\mathrm{nr}}} \frac{dt}{a(t)} \approx 30\;\Mpc \left(\frac{T_x}{T}\right)
\left(\frac{10\;\eV}{m_x}\right),
\label{eq:freestreaming_length}
\eeq
where $T$ is the temperature of the SM and $t_\mathrm{nr}$ is the time at which the 
unstable particle, $x$, becomes non-relativistic. In performing this integral, we have adopted a thermal distribution for the particle's momentum (but not necessarily for its number density) with a temperature, $T_x$.\footnote{For stable matter, the free-streaming length 
continues to grow after $t_\mathrm{nr}$, such that the 
relevant upper limit in the free-streaming integral is 
the time of matter-radiation equality~\cite{Kolb:1990vq}. 
In our case, however, 
we are interested in particles that decay at around $t_\mathrm{nr}$.} For Bose-Einstein and 
Fermi-Dirac distributions, $T_x(t_\mathrm{nr}) \approx m_x/3$.
Thus we see that the interesting parameter space 
with $T_x \sim T$ consists of particles with $m_x \sim 1 - 10\;\eV$. Lighter particles will remain 
relativistic throughout the epoch of recombination and therefore 
behave as an extra component of radiation, while heavier particles have a negligible (from the CMB perspective) free-streaming length, making them indistinguishable from decaying CDM. These conclusions are 
altered if $T_x \ll T$ or $T_x \gg T$. In such cases, however, the density of the decaying component tends (in many models) to be either negligibly low, or too high to be consistent with the successful predictions of Big Bang Nucleosynthesis (BBN). With this in mind, we will focus in this study on the case in which $T_x \sim T$ and $m_x \sim 1 - 10\;\eV$.

For particles in this mass range, which were relativistic 
before the CMB era, their initial abundance can be conveniently written in terms of $\Neffx$:
\beq 
\rho_x = \frac{7}{8}\; \left(\frac{\pi^2}{15}\right)\;\Neffx\; T_\nu^4 \;\;\;\;\;\;(T_x \gg m_x).
\label{eq:dwdm_initial_abundance}
\eeq
For example, a relativistic species that was in thermal equilibrium with the SM bath at some point in time has an energy density that corresponds to the following:
\beq
\Neffx = \frac{4}{7} \, g_x \left(\frac{T_x}{T_\nu}\right)^4,
\label{eq:dwdm_initial_abundance_from_equilibrium}
\eeq
where $g_x$ is the effective number of internal degrees-of-freedom; for a real scalar (Weyl fermion) we have $g_x = 1\;(2 \times 7/8)$. We will also 
consider cases in which equilibrium was not attained, and for which the more general expression (Eq.~\ref{eq:dwdm_initial_abundance}) should be used.

If the DWDM is already present at the time of neutrino decoupling, $T\sim \MeV$, then $\Neffx$ is constrained by the observed helium and deuterium abundances~\cite{Pitrou:2018cgg,Fields:2019pfx}.
For example, Ref.~\cite{Pitrou:2018cgg}  uses these abundances to
place an upper bound of $\Neffx < 0.4$ at the 95\% confidence level. 
Note that Eq.~\ref{eq:dwdm_initial_abundance} only specifies the initial abundance of the decaying component. The non-standard contribution (from DWDM and its 
decay products) to the energy density during the era of matter-radiation equality depends not only on this, but also on the DWDM particle mass and decay rate. Thus the non-photon radiation that the CMB and BBN are sensitive to are different, illustrating the complementarity of these two probes.

DWDM can arise in a variety of particle physics scenarios. These can be classified by 
their coupling to the SM bath, or lack thereof. 
The simplest possibility is that DWDM and its decay products are part of a completely decoupled dark sector that was 
never in thermal contact with the SM. 
Such dark sectors appear in a wide range of ultraviolet completions of the SM~\cite{Halverson:2018xge} and are essentially unconstrained, except 
through their gravitational impact on BBN and on the CMB. 
The only troubling feature of the DWDM model in this context is that in order to have a viable 
and interesting model we must have $T_x\lesssim T$ (as opposed to $T_x\ll T$ or $T_x\gg T$), 
which, in the absence of equilibrium (or at least production from the SM bath), would require something of a coincidence.

This issue is remedied in models where the dark sector is produced from the SM bath, naturally leading to $T_x\lesssim T$. 
If the coupling between the DWDM and the SM is large enough, the two sectors will be in thermal equilibrium in the early universe, maintaining $T = T_x$ up to the time of their decoupling. Simple examples of particles that can be produced from the SM bath include
sterile neutrinos~\cite{Dodelson:1993je,Shi:1998km,Abazajian:2001nj} or scalar fields that mix with the Higgs boson~\cite{Fradette:2018hhl}. Many different couplings are possible through renormalizable and non-renormalizable interactions which can be important at different times (see, for example, Ref.~\cite{Baumann:2016wac}). We will consider the sterile neutrino and Higgs portal scalar as illustrative examples, and will discuss the constraints that can be placed on their interactions below.

For a sufficiently large coupling (referred to as a mixing angle, $\theta$, in both models) the new particle equilibrates and eventually freezes out while relativistic, 
as the relevant interaction rates fall below that of Hubble expansion. 
Since relativistic species in equilibrium with the SM 
contribute $\DNeff \gsim 0.6\;(1)$ for a real scalar (Weyl fermion), BBN bounds imply that we must have $T_x < T_\nu$ (see Eq.~\ref{eq:dwdm_initial_abundance_from_equilibrium}). 
This can be easily achieved 
if the relativistic freeze-out of $x$ occurs before some of the SM particles become non-relativistic (or before a phase transition). This will have the effect of reducing the effective number of relativistic degrees-of-freedom in entropy, $g_{*S}$, as well as the value of $T_x/T_\nu$. 
Consistency with the light element abundances requires the freeze-out temperature of such a particle to be greater than $\sim 50\;\MeV$ or $\sim 150\;\MeV$ for the case of
a real scalar or Weyl fermion, respectively (corresponding to $T_x/T_\nu = 0.9$ and $0.8$). For the Higgs portal scalar, this is easily achieved due to its enhanced couplings to top quarks and other heavy SM particles. However, for this scalar to reach equilibrium with the SM requires the scalar-Higgs mixing 
angle to be $\theta\gtrsim 10^{-6}$~\cite{Fradette:2018hhl}, which is robustly excluded by stellar energy loss arguments (which require $\theta < 3\times 10^{-10}$)~\cite{Hardy:2016kme}.
In the sterile neutrino scenario, production via active-sterile oscillations peaks well below $150\;\MeV$ (in the non-resonant case the conversion rate is maximized at $\sim 30\;\MeV \;(m_x/10\;\eV)^{1/3}$~\cite{Dodelson:1993je}). 
Thus, in both of these scenarios, astrophysical bounds rule out the possibility of complete thermalization. 

Alternatively, a cosmologically relevant abundance could be accumulated through sub-Hubble processes, such as freeze-in. 
In this case we can estimate the value of $\DNeff$ relevant for BBN constraints (and for the initial conditions of our calculations, as described in the following section) 
using Eq.~\ref{eq:dwdm_initial_abundance} and the results of Refs.~\cite{Dodelson:1993je,Fradette:2018hhl}. Since 
those works were focused on the case of stable DM, we will define $\Omega_x$ as the abundance that the DWDM would have today 
{\it if it did not decay}, normalized to the critical density. 
Redshifting backwards from today to $T_x\gg m_x/3$ results in the following:
\beq
\Neffx \approx \left(\frac{40\;\eV}{m_x}\right)\Omega_x,
\eeq
where we have approximated the transition from non-relativistic to relativistic as instantaneous at $T_x = m_x/3$.
We can now estimate the contribution of DWDM to $\DNeff$ by taking the value of $\Omega_x$ predicted from Dodelson-Widrow production~\cite{Dodelson:1993je} 
or from Higgs portal freeze-in~\cite{Fradette:2018hhl}:
\beq
\Neffx \approx 0.2 \begin{cases}
  \left(\frac{m_x}{10\;\eV}\right)\left(\frac{\theta}{3\times 10^{-3}}\right)^2 & \text{sterile neutrino}\\
  \left(\frac{\theta}{2\times 10^{-7}}\right)^2 & \text{Higgs portal scalar}
\end{cases}
\eeq
where the mixing angles have been normalized to yield a value of $\DNeff$ that is compatible with BBN. Sterile neutrinos with this range of mixing angles are compatible with current laboratory constraints~\cite{deGouvea:2015euy}, especially if the mixing is predominantly with $\nu_\mu$ or $\nu_\tau$. In contrast, 
the range of Higgs portal mixing angles that are required to produce a significant energy density are still excluded by stellar energy losses~\cite{Hardy:2016kme}; 
in fact, saturating this constraint yields $\Neffx \sim 10^{-7}$, which is negligible for our purposes. 
For this reason, we focus on the sterile neutrino scenario in what follows. 
An additional simplifying feature of the Dodelson-Widrow production mechanism is that 
the momentum distribution of DWDM is nearly thermal, with the same temperature as the SM neutrinos~\cite{Dodelson:1993je}. 

For completeness, we will mention another scenario that can result in $T_x \sim T_\nu$. In this class of models, SM neutrino interactions produce particles in a dark sector (which includes DWDM) after the time of neutrino decoupling, and ultimately equilibrium between the dark sector and the SM neutrinos is attained~\cite{Chacko:2003dt,Chacko:2004cz,Berlin:2017ftj,Berlin:2019pbq}. 
Since this ``late'' equilibration scenario only transfers energy from the SM neutrinos into the dark sector, it does not alter the total energy density and thus is consistent with all constraints from BBN. The combined energy density of 
neutrinos and dark sector particles corresponds to $\Neff \approx 3$ as long as all of the particle species are relativistic; 
massive particles in the dark sector increase $\Neff$ as they become non-relativistic. DWDM is easily implemented in this framework as the lightest of such massive 
states in the dark sector (such that by the recombination era, it is the only dark sector species that is contributing to the energy density).

The last ingredient we need to address is the decay of the DWDM. In the sterile 
neutrino case, it is natural to consider decays to Majorons, $\phi$, the pseudo-Nambu-Goldstone 
bosons of spontaneous lepton number breaking~\cite{Chikashige:1980ui,Chikashige:1980qk,Schechter:1981cv}, and light neutrinos, 
i.e. $x \rightarrow \phi + \nu$.
Similar models were considered in Refs.~\cite{Bharadwaj:1997dz,Lopez:1998jt,Kaplinghat:1999xy,Lopez:1999ur,Chacko:2019nej}. In these models, the lifetime of the DWDM, $\tau_x$, is related to its mass and to the scale of 
spontaneous lepton number breaking, $f$:
\beq
\tau_x \sim \frac{16\pi f^2}{m_x^3} \simeq  10^5\;\mathrm{yr}\;\left(\frac{f}{3\times 10^5\;\GeV}\right)^2\left(\frac{10\;\eV}{m_x}\right)^3,
\label{eq:sterile_decay}
\eeq
where we have normalized the lifetime to the age of the universe just prior to recombination.
The new scalars (the Majoron and its CP-even partner) couple to SM neutrinos with a strength, $m_\nu/f$, which 
is easily compatible with laboratory bounds from searches for rare $\tau$ and meson decays~\cite{Blum:2014ewa,Berryman:2018ogk,Kelly:2019wow,Krnjaic:2019rsv}, and 
neutrinoless double $\beta$ decay~\cite{Agostini:2015nwa, Blum:2018ljv, Brune:2018sab} (see Ref.~\cite{Blinov:2019gcj} for a recent compilation of these constraints).
Couplings of these scalars to charged leptons are also generated 
at one loop, but the resulting rates of lepton flavor violation are currently unobservable~\cite{Garcia-Cely:2017oco}.

Finally, we note that the Majoron can be radiated in any process involving the neutrinos. However, because of 
its small coupling and (possibly vanishing) mass, its freeze-in yield is cosmologically irrelevant.\footnote{The Majoron could instead be produced 
in the ultraviolet, e.g., when lepton number is spontaneously broken or during inflation. If this occurs at a high enough 
temperature, however, its energy density will be diluted through SM entropy injections as described above.  
If the Majoron mass is near the eV scale, it can acquire a significant abundance 
through inverse decays, $\nu\nu\rightarrow \phi$~\cite{Chacko:2003dt,Chacko:2004cz,Berlin:2017ftj,Berlin:2019pbq}. 
For these masses, however, the Majoron will not act as dark radiation throughout the epoch of recombination, so we do not consider this possibility further.
}
In the absence of an initial Majoron population, inverse decays $\phi + \nu \rightarrow x$ 
are unimportant. However, as the $x$ population decays, it can produce a significant bath of $\phi$ and $\nu$ that 
can backreact via this process. We will limit ourselves to the regime where the $x$ 
decays occur after they have become non-relativistic, allowing us to neglect this backreaction. This 
requirement constrains the lifetime and mass of the DWDM particle by enforcing $T_{x,\mathrm{dec}} < T_{x,\mathrm{nr}} \approx m_x/3$,
where $T_{x,\mathrm{dec}}$ is the temperature of the DWDM bath at the time of its decay. 
We find that the $x$ population decays after becoming non-relativistic if the following condition is met:
\beq
m_x\gtrsim 2\;\eV 
\begin{cases}
  \left(\frac{10^5\;\yr}{\tau_x}\right)^{1/2} & \tau_x \ll 10^5\;\yr \\
  \left(\frac{10^5\;\yr}{\tau_x}\right)^{2/3} & \tau_x \gg 10^5\;\yr
\end{cases}
\label{eq:nr_decay_condition}
\eeq
where we have assumed that $T_x = T_\nu$, as motivated 
by the Dodelson-Widrow production mechanism. These two cases 
approximately correspond to decays that take place during radiation or matter domination, respectively.
This condition will enable us to make an important simplification to the Boltzmann equations in the following section~\cite{Lopez:1999ur}. 
Since particles with different momenta decay at different times, the backreaction rate can be different for different 
regions of phase space, so Eq.~\ref{eq:nr_decay_condition} is only a rough guideline. In the following 
sections, we limit our discussion to the case in which $m_x \gtrsim \;\eV$.

\section{Boltzmann Evolution}
\label{sec:boltzmann}

In this section, we derive the Boltzmann equations and initial conditions necessary to determine the cosmological impact of DWDM and its dark radiation decay products.

\subsection{Background Equations}

\begin{figure}
    \centering
    \includegraphics[width=0.47\textwidth]{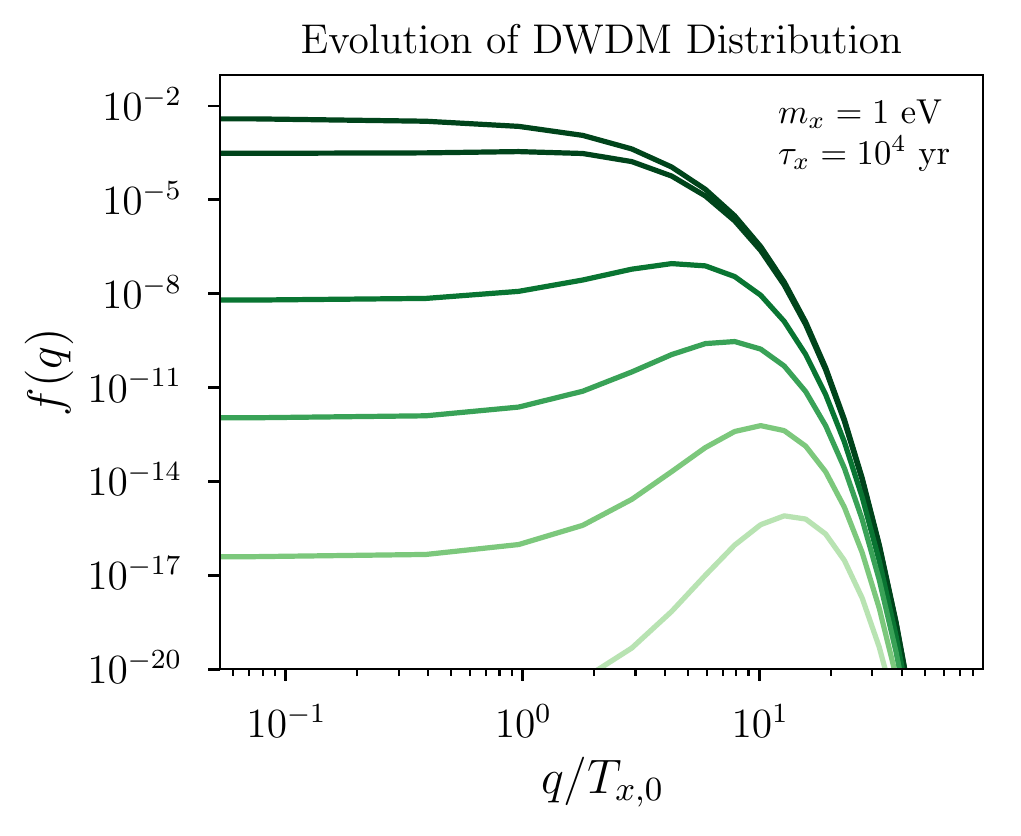}
    \includegraphics[width=0.47\textwidth]{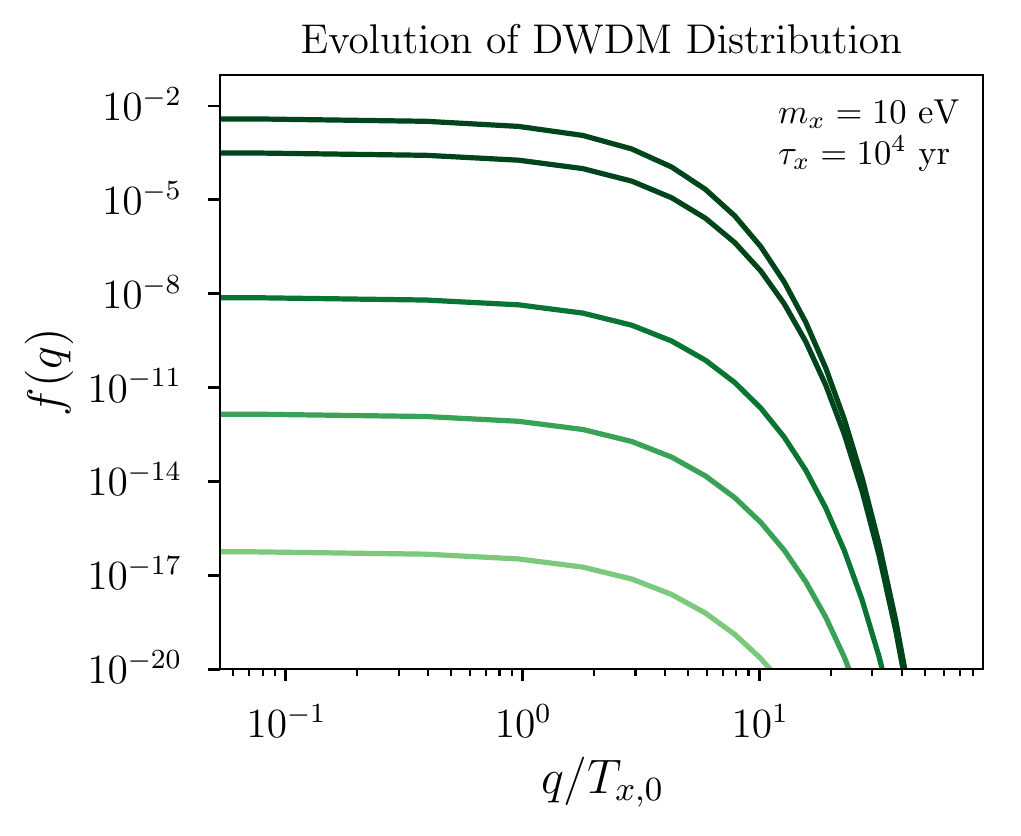}
    \caption{Evolution of the decaying warm dark matter (DWDM) background phase space distribution, $f$, for $m_x=1\;\eV$ (left panel) and $m_x=10\;\eV$ (right panel). The different colored lines correspond to different times, with darker colors representing earlier times. If the mass is smaller than the typical momentum at the time of the decay, slower-moving particles decay first. If the DWDM is already non-relativistic, however, the decays simply rescale the distribution while preserving its shape. 
    The conformal momentum on the horizontal axis is normalized to the temperature DWDM would have today if it did not decay.}
    \label{fig:distribution_evolution}
\end{figure}

As a result of the non-negligible velocities of the DWDM, many of these particles will have significant boost factors and thus decay later than if they had been at rest. Moreover, the DWDM mass will become 
comparable to the typical particle momentum, leading to a change in the equation-of-state parameter.
These facts mean that we cannot use the standard phase-space integrated equations for the 
energy or number density of DWDM particles. Instead, we solve the full momentum-dependent Boltzmann equation for the phase space distribution, $f$~\cite{Kaplinghat:1999xy}: 
\beq
  \pd{f}{t} - H \frac{p^2}{E}\pd{f}{E} = a^{-1}\pd{f}{\tau} = -\frac{1}{E} m_x \Gamma_x f,
  \label{eq:dwdm_boltzmann}
\eeq
where in the first equality we have taken $f$ to be a function of the conformal momentum, $q = a p$, and the conformal time, $\tau$. 
The collision term on the right-hand side implements DWDM decays into dark radiation with a decay rate, $\Gamma_x$. For simplicity, we have neglected final state dark radiation Pauli-blocking and Bose-enhancement factors. It will be useful to define the conformal-time collision term:
\beq
\left(\pd{f}{\tau}\right)_C = - \frac{a^2}{\epsilon} m_x \Gamma_x f,
\label{eq:dwdm_collision_term}
\eeq
where $\epsilon = \sqrt{q^2 + a^2 m_x^2}$.\footnote{This form of the collision term is also valid in the \emph{perturbed} 
universe in the synchronous gauge. In other gauges (such as the conformal Newtonian gauge), there are additional factors of metric perturbations 
necessary to convert from proper time, $t$, to conformal time, $\tau$.} In Fig.~\ref{fig:distribution_evolution}, we 
show the solution to Eq.~\ref{eq:dwdm_boltzmann} for two different DWDM masses. The $m_x=1\;\eV$ case 
shown in the left panel clearly demonstrates the fact that the ``slower'' parts of the distribution 
decay first. In the right panel, we show the $m_x=10\;\eV$ case, for which the decays occur when the DWDM is sufficiently non-relativistic 
that only the overall normalization of the distribution is significantly impacted.

The background density and pressure are determined by integrating $f$:
\begin{align}
  \rho & = a^{-4} \int d^3 q \epsilon f(q) \\
  p & = a^{-4} \int d^3 q \frac{q^2}{3\epsilon}f(q).
\end{align}
Performing the momentum integration yields an equation for the total DWDM energy density, $\rho$:
\beq
\dot \rho + 3 \Hcon (\rho  + p)  = - a \Gamma_x m_x n,
\label{eq:dwdm_bg_eq}
\eeq
where $p$ is the pressure, $\Hcon = a H$ is the conformal Hubble rate and the dot 
denotes a derivative with respect to conformal time. 
It is important to note that $m_x n$ appears in the collision term and 
not $\rho$; this distinction is important for semi-relativistic decays.

The DWDM decays into dark radiation, which has a constant equation of state, so its background evolution is 
fully specified by the integrated equation:
\beq  
\dot{\rho}_{\dr} + 4 \Hcon \rdr = + a \Gamma_x m_x n,
\eeq
where the source term follows from Eq.~\ref{eq:dwdm_bg_eq} and the first law of thermodynamics. 
Since we will derive the Boltzmann equations for the perturbed dark radiation density, we also 
note that the collision term for the momentum-dependent equation is 
\begin{align}
\left(\pd{f_\dr}{\tau}\right)_C (q_1)& = + 2 \times \frac{a}{2E_1} \int d\Pi_2 d\Pi_3 |\mathcal{M}|^2 (2\pi)^4 \delta^4(p_3 - p_1 - p_2) f(p_3) \nonumber \\
& = +  2\times a(16\pi m_x \Gamma_x) I(p_1), 
\label{eq:dr_collision_term_full}
\end{align}
where $p_1$ ($q_1$) is the physical (comoving) momentum of one of the dark radiation particles, $p_2$ is the momentum of the other dark radiation particle 
produced in the decay of a particle of mass $m_x$, and momentum $p_3$.
In the above we replaced $|\mathcal{M}|^2$ by $16\pi\Gamma_x$ (the coefficient depends on whether the final state consists of identical particles), 
and the explicit factor of two captures one of two effects. First, if final state particles are different (but still massless), then there are 
two collision terms corresponding to each particle type populating the region of phase space around $p_1$, which can be massaged into the same form. 
Second, if the two particles are identical, then $|\mathcal{M}|^2$ should be replaced by $32\pi\Gamma_x$ instead. In Sec.~\ref{sec:dr_perturb} we will 
simplify the integral
\beq
\label{eq:dr_collision_int0}
I(p_1) = \frac{1}{2E_1} \int d\Pi_2 d\Pi_3 (2\pi)^4 \delta^4(p_3 - p_1 - p_2) f(p_3)
\eeq
to derive the perturbed dark radiation collision terms.

\begin{figure}
    \centering
    \includegraphics[width=0.47\textwidth]{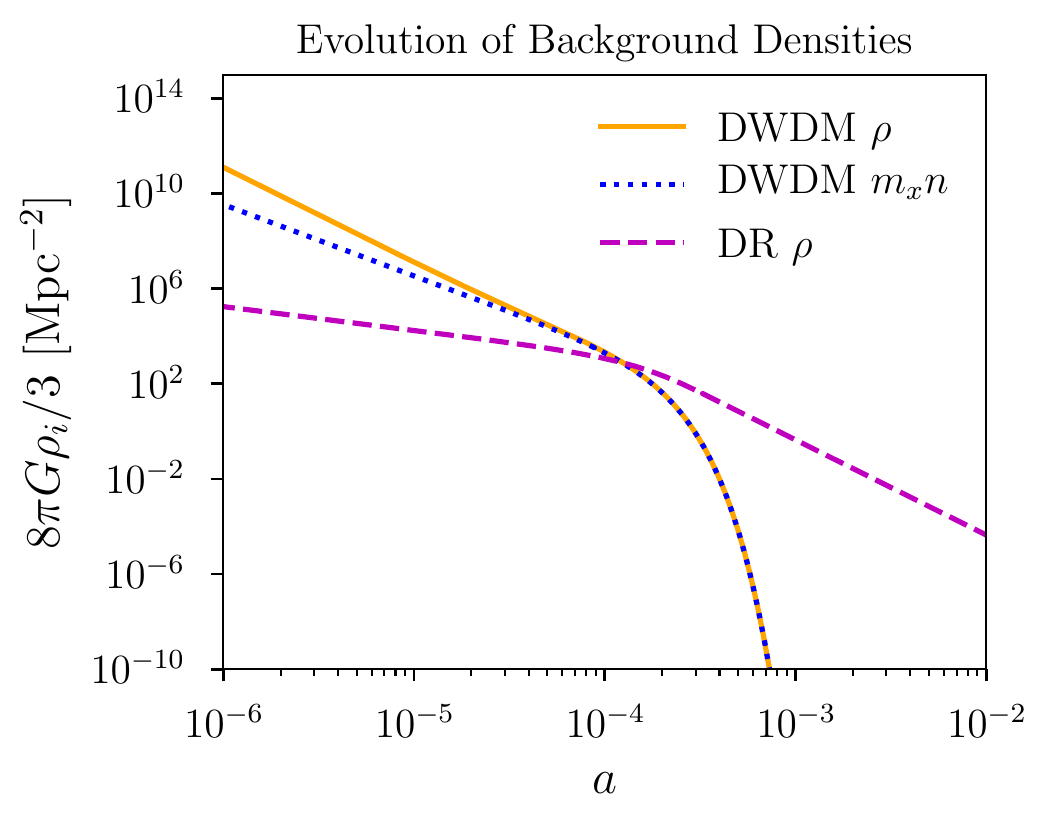}
    \caption{Evolution of the background densities of decaying warm dark matter (DWDM) and dark radiation (DR), as a function of the scale factor for $m_x = 10\;\eV$ and $\tau_x \approx 10^4$ years. At early times, the DWDM energy density redshifts like radiation. When $T_x \sim m_x$, it transitions to matter-like dilution with $\rho$ tracking the number density, $n$.
      Dark radiation does not evolve as $a^{-4}$ initially due to the energy injected from DWDM decays.}
    \label{fig:bg_evolution}
\end{figure}

In Fig.~\ref{fig:bg_evolution} we show the evolution of the background number and energy densities of DWDM and dark radiation. 
The DWDM energy density redshifts as radiation until its temperature becomes comparable to its mass. At this point, 
$\rho$ approaches the non-relativistic expectation, $m_x n$. The density becomes exponentially suppressed when $\Gamma_x n/(H \rho) \sim 1$.
Note that the dark radiation density does not initially redshift as $a^{-4}$ because of energy injection from DWDM decays.

\subsection{Perturbation Equations for DWDM}
\label{sec:DWDM_perturb}
In this section, we derive the Boltzmann equations for the DWDM fluid in an inhomogeneous universe.
We start with the perturbed Friedmann-Robertson-Walker spacetime in synchronous gauge~\cite{Ma:1995ey,Hu:2004xd}: 
\beq
ds^2 = a^2(\tau) \left[ -d\tau^2 + (\delta_{ij} + 2h_{ij}) dx^i dx^j\right],
\eeq
where $\tau$ is the conformal time, and $h_{ij} = h \hat{k}_i \hat{k}_j/2 + 3\eta(\hat{k}_i \hat{k}_j - \delta_{ij}/3)$ is the metric perturbation in Fourier space. 
Following standard conventions~\cite{Ma:1995ey}, we write
\beq
f = f^{(0)}(q) \left[1 + \Psi (q, n_i, x_i, \tau)\right],
\eeq
where $f^{(0)}$ is the background distribution (a solution of Eq.~\ref{eq:dwdm_boltzmann} with $f=f^{(0)}$), $\Psi$ is the position and momentum-dependent perturbation, and $n_i$ is the 
direction of $q$.
The Boltzmann equation for $\Psi$ in Fourier space takes the form~\cite{Ma:1995ey}: 
\beq
\pd{\Psi}{\tau} + i\,\frac{q}{\epsilon}
	(\vec{k}\cdot \hat{n})\Psi +
    \frac{d\ln f^{(0)}}{d\ln q}\, \left[\dot{\eta} - \frac{\dot{h}+6\dot{\eta}}{2}(\hat{k}\cdot\hat{n})^2 \right] =
    \frac{1}{f^{(0)}}\,\left(\pd{f}{\tau}\right)_C^{(1)},
\eeq
where
\beq
\left(\pd{f}{\tau}\right)_C^{(1)} = \left(\pd{f}{\tau}\right)_C - (1+\Psi)  \left(\pd{f^{(0)}}{\tau}\right) = 0,
\eeq
and $(\partial f/\partial \tau)_C$ is given in Eq.~\ref{eq:dwdm_collision_term}. The last equality in this expression follows 
from Eq.~\ref{eq:dwdm_boltzmann} and the fact that the collision term is linear in the full distribution, $f$. 
This cancellation is a special feature of the synchronous gauge (see the footnote following Eq.~\ref{eq:dwdm_collision_term}).

Note that the dependence on the direction of $q$, $\hat n$, is only through $\vec{k}\cdot \hat{n}$. If the same is 
true of the collision term, then the solution has azimuthal symmetry about $\vec{k}$, enabling an expansion in Legendre polynomials, $P_l$.
With this assumption, we expand as follows:
\beq
\label{eq:DWDM_legendre_decomp}
\Psi(\vec{k},\hat{n},q,\tau)
	= \sum_{l=0}^\infty (-i)^l(2l+1) \Psi_l(\vec{k},q,\tau)
	P_l(\hat{k}\cdot\hat{n}),
\eeq
which leads to the following Boltzmann hierarchy in synchronous gauge:
\begin{subequations}
\begin{align}
\dot{\Psi}_0 &= -\frac{qk}{\epsilon}\Psi_1 +\frac{1}{6}\dot{h} \frac{d\ln f^{(0)}}{d\ln q} \,, \\
\dot{\Psi}_1 &= \frac{qk}{3\epsilon} \left(\Psi_0 - 2 \Psi_2 \right) \,,\\
\dot{\Psi}_2 &= \frac{qk}{5\epsilon} \left(2\Psi_1 - 3\Psi_3 \right) - \left( \frac{1}{15}\dot{h} + \frac{2}{5} \dot{\eta} \right)
  \frac{d\ln f^{(0)}}{d\ln q} \,,\\
\dot{\Psi}_l &= \frac{qk}{(2l+1)\epsilon} \left[ l\Psi_{l-1} - (l+1)\Psi_{l+1} \right]\,, \;\;\;l \geq 3 \,.
\end{align}
\label{eq:dwdm_hierarchy}
\end{subequations}
These equations are identical to the massive neutrino case discussed in Ref.~\cite{Ma:1995ey}, except now the 
terms proportional to $d\ln f^{(0)}/d\ln q$ are time-dependent. Note that there are no new terms that are proportional to $\Gamma_x$; 
this is special to the synchronous gauge (this was also noted in Ref.~\cite{Kaplinghat:1999xy}). 
The physical quantities that source Einstein's equations (density, pressure, velocity and shear perturbations) 
are obtained by integrating the multipole coefficients as described in Ref.~\cite{Ma:1995ey}.
We implemented Eq.~\ref{eq:dwdm_hierarchy} in \CLASS in analogy to the massive neutrino module~\cite{2011JCAP...09..032L}.

\subsection{Perturbation Equations for Dark Radiation}
\label{sec:dr_perturb}
In order to derive the collision term for the perturbed dark radiation Boltzmann equations, we will simplify 
the full expression of Eq.~\ref{eq:dr_collision_term_full}. Because the dark radiation equation of state 
is constant, the Boltzmann equations can be expressed in terms of the 
momentum-integrated perturbations, $F_{\dr}$, in analogy to massless neutrinos~\cite{Ma:1995ey}:
\beq
F_\dr(\vec{k}, \hat{n}_1, \tau) = \frac{\int dp_1 p_1^3 f_{\dr}^{(0)}(p_1) \Psi_{\dr}}{\int dp_1 p_1^3 f_{\dr}^{(0)}(p_1)} r_\dr,
\eeq
where $\hat n_1$ is the direction of the dark radiation three-momentum, $\vec{p}_1$, and 
$r_\dr = \rho_\dr a^4/\rho_c$~\cite{Poulin:2016nat} (the normalization of $r_\dr$ is a arbitrary; 
  we are following \CLASS conventions, but, e.g., Ref.~\cite{Kaplinghat:1999xy} normalizes $\rho_{\dr}$ to $\rho_\nu$).
The Boltzmann equations can be cast completely in terms of $F_\dr$, so the precise form the dark radiation distribution, $f_\dr^{(0)}$, is not important. 
The collision term for the $F_\dr$ variable will therefore involve the integral 
\beq
I_F(\hat{n}_1) = \int dp_1 p_1^3 I(p_1) \equiv I_F^{(0)} + I_F^{(1)},
\label{eq:dr_collision_int1}
\eeq
where $I(p_1)$ is given in Eq.~\ref{eq:dr_collision_int0} and the last step separates the background and perturbed contributions. We already have a background evolution equation for dark radiation, but we will use this opportunity to check whether we find the same result. The DWDM background distribution, $f^{(0)}$, only depends on 
the magnitude of the DWDM momentum, so the angular integrals can be carried out explicitly (see Appendix~\ref{sec:dr_source_terms}):
\beq
I_F^{(0)} = \frac{\pi}{16} n.
\eeq
Since $\rho_{\dr} = \int dp_1 p_1^3 f_{\dr}/(2\pi^2)$, using Eqs.~\ref{eq:dr_collision_term_full} and~\ref{eq:dr_collision_int1}, 
we find that
\beq
\dot \rho_{\dr}\supset a (32\pi m_x \Gamma_x) I_F^{(0)}/(2\pi^2) = + a m_x \Gamma_x n, 
\eeq
in agreement with expectations. 

As we show in Appendix~\ref{sec:dr_source_terms} the dark radiation perturbation term evaluates to 
\beq
  \label{eq:dr_collision_int3}
  I_F^{(1)}  = \sum_{l=0}^{\infty} (-i)^l (2l+1) P_l(\hat k\cdot \hat n_1) \times \frac{1}{32\pi} \int dp_3 p_3^2 f^{(0)}(q_3)\Psi_l(q_3) \mathcal{F}_l(p_3/E_3),
\eeq
where
\beq
\mathcal{F}_l(x) = \frac{(1-x^2)^2}{2}\int_{-1}^{+1} \frac{du P_l(u)}{(1 - xu)^3}.
\eeq
Each term in the sum above sources a single moment in the dark radiation Boltzmann hierarchy (the Legendre expansion of the $F_\dr$):
\beq
F_{\dr}(k, q_1, \hat n_1, \tau) = \sum_{l=0}^\infty (-i)^l(2l+1) F_{\dr,l}(k,q_1,\tau)
P_l(\hat{k}\cdot\hat{n}_1).
\eeq

The first three moments are related to the energy density, velocity and shear perturbations of the dark radiation fluid~\cite{Poulin:2016nat}:
\beq
\label{eq:dr_moment_perturb_relation}
F_{\dr,0} = \delta_\dr r_\dr , \;\;\; F_{\dr,1} = \frac{4\theta_\dr}{3k} r_\dr, \;\;\; F_{\dr,2} = 2\sigma_{\dr} r_\dr.
\eeq

The collision terms for each component, $F_{\dr,l}$, can be read off from Eqs.~\ref{eq:dr_collision_int3} and \ref{eq:dr_collision_term_full}:
\beq
(\dot{F}_{\mathrm{dr},l})_C \equiv r_{\dr}\frac{\int d p_1 p_1^3 \left(\pd{f_\dr}{\tau}\right)^{(1)}_C}{\int d p_1 p_1^3 f^{(0)}_\dr(p_1)} = 
  \dot{r}_{\dr} \frac{\int dq q^2 f^{(0)}(q)\Psi_l(q) \mathcal{F}_l(q/\epsilon)}{\int d q q^2 f^{(0)}(q)},
  \eeq
where 
\beq
\dot{r}_{\dr} = \frac{1}{\rho_c}\frac{d\rho_\dr a^4}{d\tau} = r_{\dr} \frac{a m_x \Gamma_x n}{\rho_\dr}.
\eeq
The first few functions, $\mathcal{F}_l(x)$, are shown in Table~\ref{tab:ffuncs}.
Note that the $l=0$ collision term is explicitly proportional to $\delta n/n$, 
so that the perturbation in the dark radiation energy density is proportional to the perturbation in the 
DWDM number density, as naively expected.

\begin{table}
\centering
  \begin{tabular}{|c|c|}
\hline
 $l$ & $\mathcal{F}_l(x)$ \\ 
 \hline
 0 & $1$ \\
 1 & $x$ \\
 2 & $\frac{x \left(5 x^2-3\right)+3 \left(x^2-1\right)^2 \tanh ^{-1}(x)}{2 x^3}$ \\
 3 & $\frac{-8 x^5+25 x^3+15 \left(x^2-1\right)^2 \tanh ^{-1}(x)-15 x}{2 x^4}$ \\
 4 & $-\frac{81 x^5-190 x^3+15 \left(x^2-7\right) \left(x^2-1\right)^2 \tanh ^{-1}(x)+105 x}{4 x^5}$\\
\hline
\end{tabular}
\caption{First few expressions for the function, $\mathcal{F}_l(x)$, that enters the dark radiation collision terms.\label{tab:ffuncs}}
\end{table}

Putting everything together, the dark radiation Boltzmann hierarchy in the synchronous gauge is given by:
\begin{subequations}
\begin{align}
\dot F_{\dr,0} & =  -kF_{{\dr},1}-\frac{2}{3}r_{\dr}\dot h + (\dot{F}_{\dr,0})_C ~,\\
\dot F_{\dr,1} & =  \frac{k}{3}F_{{\dr},0}-\frac{2k}{3}F_{{\rm dr},2}+(\dot{F}_{\dr,1})_C~,\\
\dot F_{\dr,2} & =  \frac{2k}{5}F_{{\dr},1}-\frac{3k}{5}F_{{\rm dr},3}+\frac{4}{15}r_{\dr}\left(\dot h + 6\dot \eta\right) + (\dot{F}_{\dr,2})_C ~,\\
\dot F_{\dr,\ell} & = \frac{k}{2\ell+1}\big(\ell F_{\dr,\ell-1}-(\ell+1)F_{\dr,\ell+1}\big) +  (\dot{F}_{\dr,l})_C \qquad \ell>2.
\end{align}
\label{eq:dr_boltzmann_hierarchy}
\end{subequations}
We have implemented these equations into \CLASS, with the collision terms computed at each time step using Gauss-Laguerre quadrature.
This is computationally costly and some approximations to speed up the calculation are described in the following section.

We note that compared to the decaying cold dark matter (DCDM) case studied in Ref.~\cite{Poulin:2016nat}, only the collision terms are different. 
As a check of this result, one can show in the limit in which the DWDM is cold and $\delta n / n = \delta\rho/\rho$ (by making use of Eq.~\ref{eq:F_limiting_form}),
\begin{align}
(\dot{F}_{\dr,0})_C & \rightarrow \dot r_\dr \delta \\ 
(\dot{F}_{\dr,1})_C & \rightarrow \dot r_\dr \theta/k,
\end{align}
while higher $l$ collision terms are suppressed by $(T_x/m_x)^{l-1}$, if the distributions 
can be taken to be thermal.
These limiting forms reproduce the DCDM collision terms in Ref.~\cite{Poulin:2016nat}. 
We can also compare our result to that of Ref.~\cite{Kaplinghat:1999xy}, which only gives a partial expression for $(\dot{F}_{\mathrm{dr},0})_C$, expanded in $q/am_x$ under the integral. Our collision term
agrees with theirs if one neglects their ``higher order'' terms, but this expansion is not always valid for DWDM, as $q/am_x$ is not necessarily negligible.

\subsection{Initial Conditions}
The starting time for the mode evolution should be well before they have entered the horizon ($k\tau \ll 1$). Following Ref.~\cite{Ma:1995ey}, one can solve the perturbation equations, order by 
order in $k\tau$ assuming that: 1) the time is early enough that both neutrinos and DWDM can be safely treated as effectively massless, and
2) the time is early enough that decays are not important, which implies that $\rho_x/\rho_\gamma$ is constant and $\rdr/\rho_x\ll 1$. 
Under these assumptions, we find that the initial conditions for the photon, baryon, neutrino, (stable) cold dark matter, and 
metric potentials are identical to those in the \lcdm case~\cite{Ma:1995ey} with $\rho_\nu \rightarrow \rho_\nu + \rho_x$ (i.e., DWDM is free-streaming 
at early times and contributes to the anisotropic stress like SM neutrinos). The initial conditions 
of DWDM and dark radiation perturbations are identical to those of massless neutrinos.

\subsection{Approximations and Sample Solutions}
We have implemented the Boltzmann equations described in the previous sections into \CLASS. 
Numerical integration of momentum-dependent equations and the integration of the solutions at each time step (in order to compute the
dark radiation collision terms) is computationally expensive, so it is worthwhile to seek approximations. 
First we limit ourselves to using 20 Gauss-Laguerre quadrature points for the evolution of the background DWDM distribution, $f^{(0)}(q)$. 
As shown in Fig.~\ref{fig:distribution_evolution}, this gives a dense enough $q$ ``grid'' to track the distributions 
for the range of masses and decay rates that we are interested in. 
The Legendre decomposition of the perturbed DWDM distribution, $\Psi(q)$, is a slowly varying function of $q$, 
so we use only $5$ quadrature points. We have checked that doubling these numbers leads to changes in the predicted $C_\ell$ values that are smaller than one percent across a wide range of parameter space.

The dark radiation collision terms require an integration of the DWDM distribution at every time step. We therefore only include 
collision terms for $l \leq 3$ in the hierarchy of Eq.~\ref{eq:dr_boltzmann_hierarchy}. Evaluating twice as many 
collision terms has negligible impact on the predicted $C_\ell$ values.

The $q$-dependence of the DWDM perturbation results in a large number of equations that need to be solved. 
Ref.~\cite{2011JCAP...09..032L} describes an approximation scheme for massive neutrinos in which the Boltzmann 
hierarchy (equivalent to Eq.~\ref{eq:dwdm_hierarchy}) is replaced by an effective fluid description (with only three equations) 
whenever a mode is deep inside the horizon ($ k \tau \gg 1$). We use the same scheme, 
employing the fluid approximation for $k\tau \geq 32$; variations of up to $\pm 50\%$ in this threshold do not appreciably 
change the predicted $C_\ell$ values.

\begin{figure}
    \centering
    \includegraphics[width=0.47\textwidth]{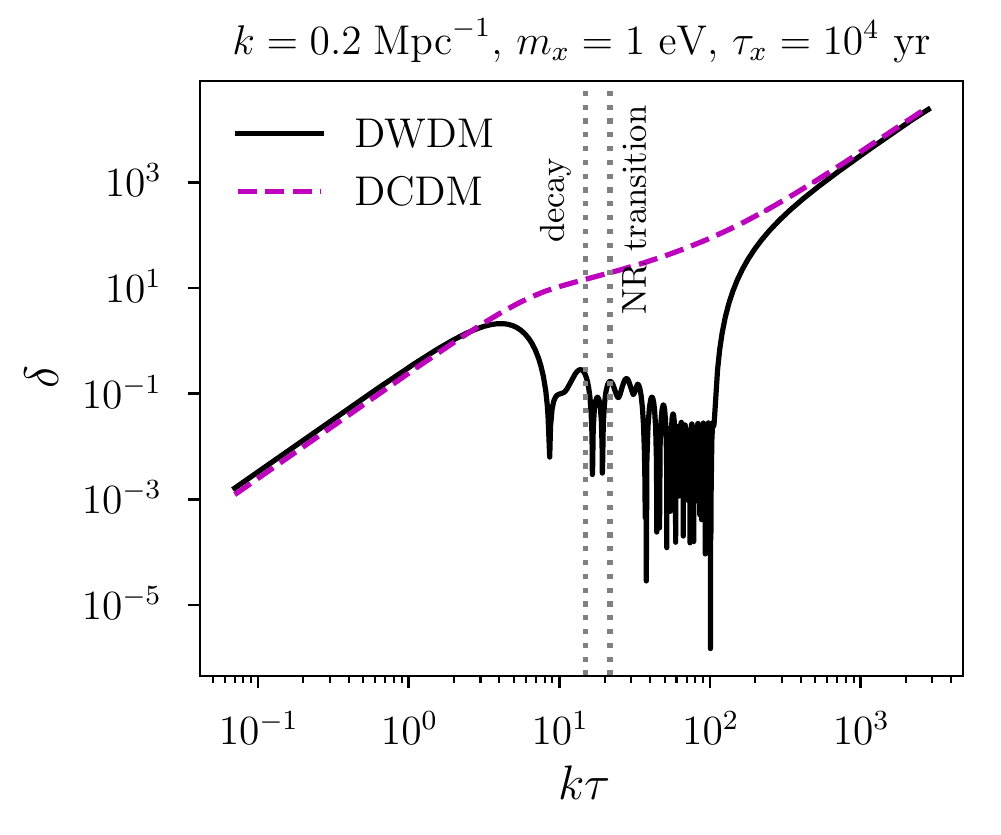}
    \includegraphics[width=0.47\textwidth]{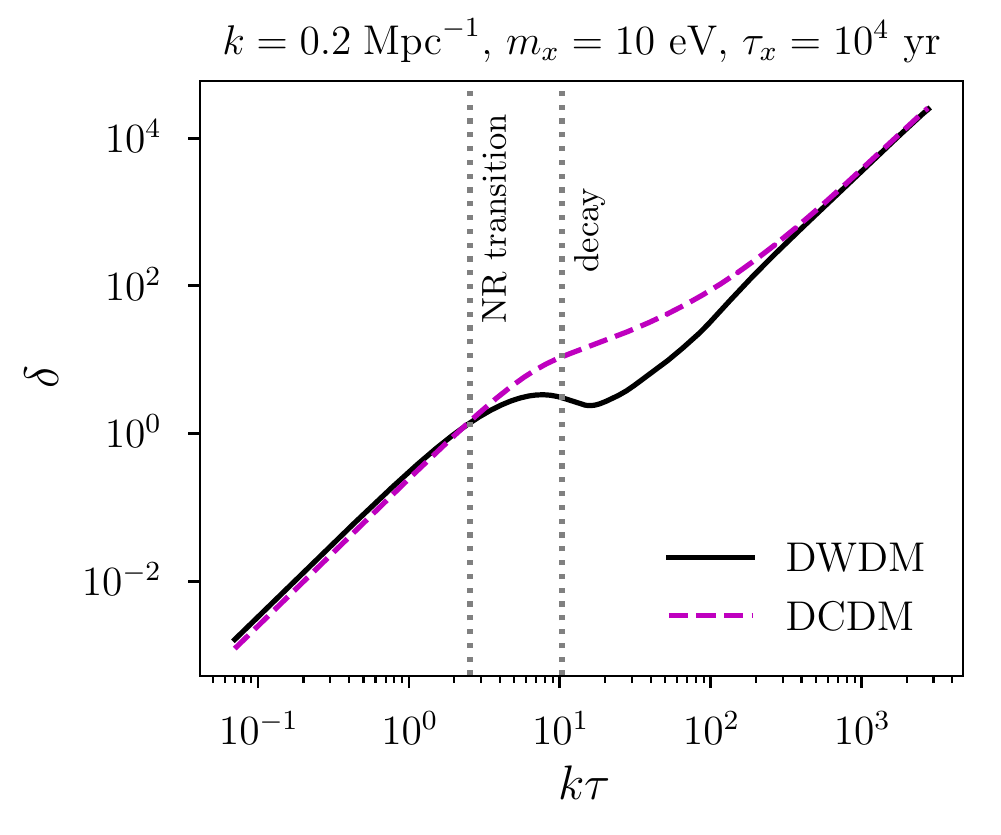}
    \caption{Evolution of the $k=0.2/\Mpc$ decaying warm dark matter (DWDM) density contrast in synchronous gauge (solid black line) compared with the decaying {\it cold} dark matter (DCDM) density contrast (dashed magenta line) for $m_x=1\;\eV$ (left panel) and $m_x=10\;\eV$ (right panel). In each panel, we have adopted a lifetime of $\tau \approx 10^4$ years for the DWDM and DCDM fluids. The vertical dotted lines denote 
    the DWDM non-relativistic transition (when $m_x\approx T_x/3$) and the decay time (when $\rho_{\mathrm{dwdm}} = \rho_\dr$). 
  }
    \label{fig:mode_evolution}
\end{figure}

In Fig.~\ref{fig:mode_evolution}, we combine the above approximations and show the evolution 
of the DWDM density contrast for $k=0.2/\Mpc$ (roughly corresponding to the high-$\ell$ part 
of the power spectrum measured by Planck), $\tau_x = 10^4$ yr, and $m_x=1$ or $10$ eV. 
We compare these results to those of a decaying {\it cold} dark matter (DCDM) model~\cite{Poulin:2016nat} with the same lifetime and that yields the same dark radiation density at late times. Depending on the mass of the DWDM particle, the decay can occur around or after the 
non-relativistic transition. In the $m_x=1$ eV case shown in the left panel, the decays are already important 
in the semi-relativistic regime. Here the density contrast is significantly suppressed by DWDM free-streaming 
compared to the same mode of DCDM~\cite{Lesgourgues:2006nd}. 
This effect is less pronounced in the $m_x=10$ eV case (right panel); however, it is still significant 
despite the fact that the decay occurs after the non-relativistic transition.
As the mass of the DWDM particle is increased further, the evolution approaches that of DCDM case (e.g., $m_x=40\;\eV$ is nearly indistinguishable from DCDM). 
Similarly, modes that enter the horizon after the non-relativistic transition behave as DCDM.

\begin{figure}
    \centering
    \includegraphics[width=0.47\textwidth]{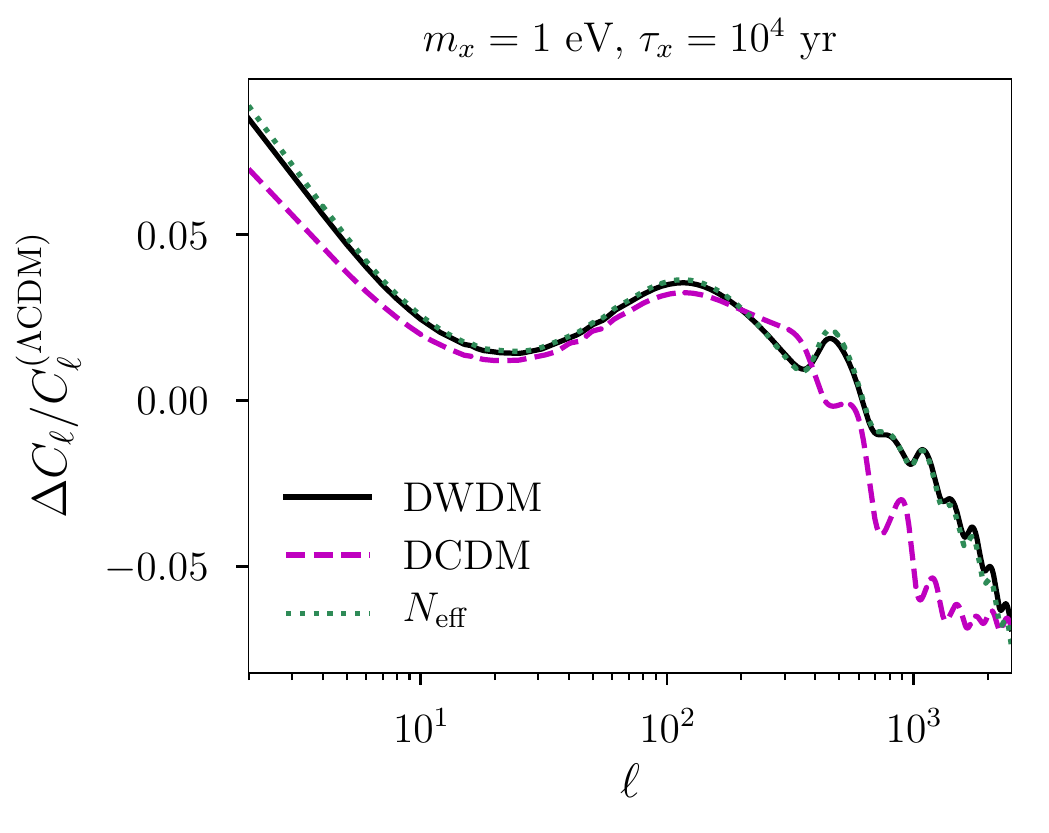}
    \includegraphics[width=0.47\textwidth]{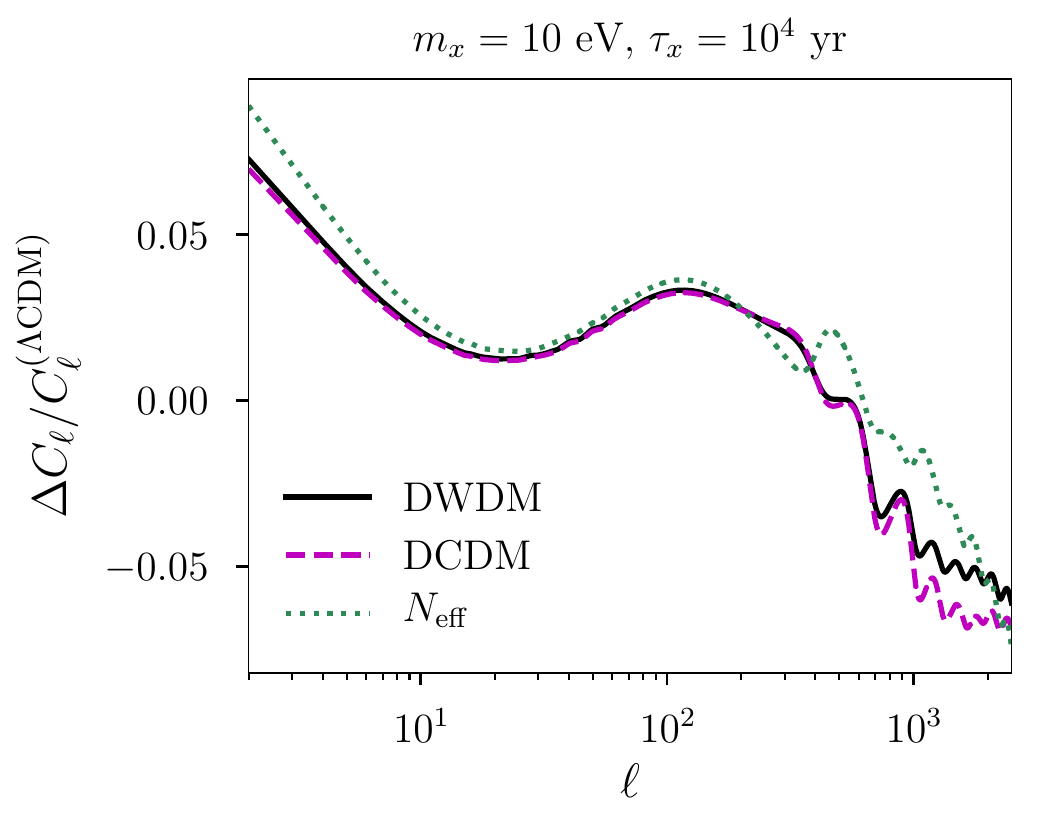}\\
    \includegraphics[width=0.47\textwidth]{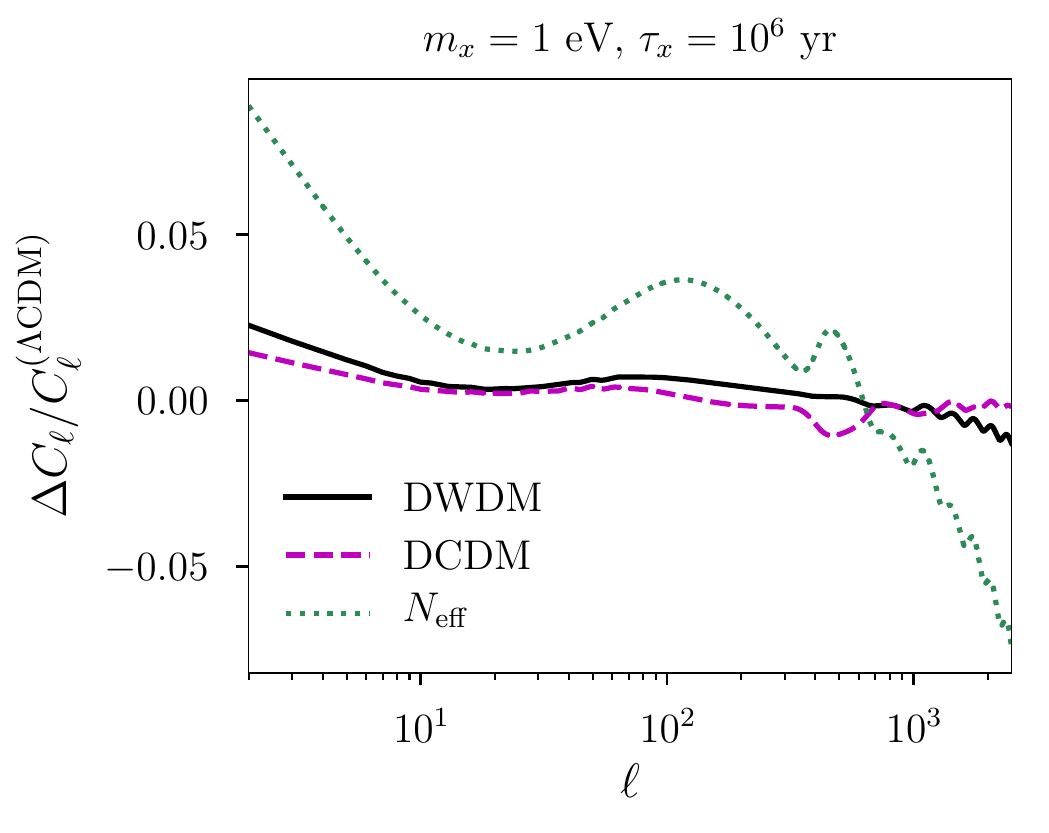}
    \includegraphics[width=0.47\textwidth]{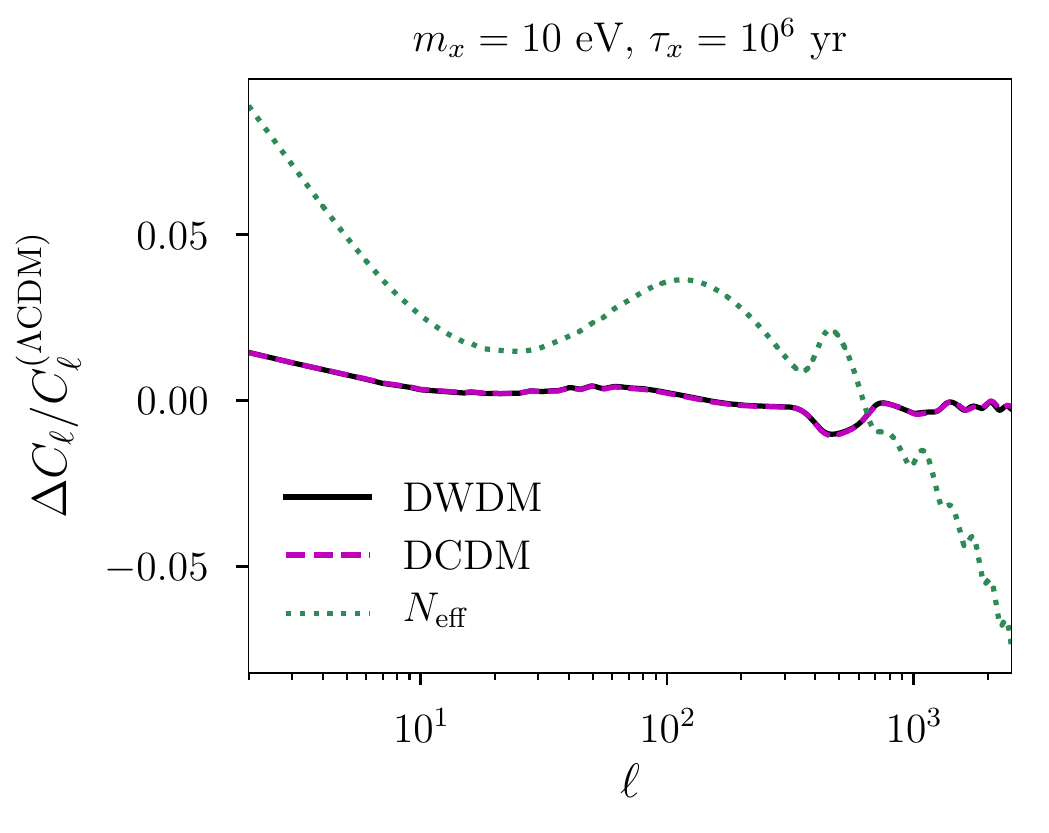}
    \caption{The fractional difference of several decaying warm dark matter (DWDM) models (solid black lines) with respect to \lcdm in the temperature angular power spectrum. We also show the same quantity for decaying cold dark matter (DCDM, dashed magenta), and for a model with additional dark radiation, parametrized by $\Neff$ (dotted green). In the left (right) column, the DWDM mass is $m_x=1\;(10)\;\eV$, while the upper (lower) row corresponds 
    to $\tau_x=10^4\;(10^{6})$ years (the same lifetimes are chosen for DCDM). For each model, the parameters are have been chosen to yield the same non-photon radiation density at late times (equivalent to $\DNeff = 0.5$).}
    \label{fig:cl_diff}
\end{figure}

In Fig.~\ref{fig:cl_diff}, we compare the temperature angular power spectrum of several models to \lcdm for fixed standard cosmological parameters (including $\theta_s$). In the left (right) column the DWDM mass is $m_x=1\;(10)\;\eV$, while the upper (lower) row corresponds to a lifetime of $\tau_x=10^4\;(10^{6})$ years. We also show the fractional shifts in $C_\ell$ that are predicted for DCDM, and in a model with dark radiation (parametrized in terms of $\Neff$). In each case, we have tuned the model parameters to give the same amount of dark radiation (equivalent to $\DNeff = 0.5$) at late times. In the upper left panel, we show that for very light DWDM and fast decay rates, this scenario is very similar to $\Neff$, as there is only a brief period of non-radiation-like evolution 
during the decay. Moreover, despite having the same decay rate, the decays happen later in the DWDM model compared to DCDM due to the semi-relativistic nature of DWDM particles. As we increase the DWDM mass to $10\;\eV$, the DWDM model interpolates between DCDM and $\Neff$, as expected. 
In the second row, we consider a longer lifetime for DWDM and DCDM. In this regime, the decays occur after recombination, at which point the effects of energy injection due to DWDM or DCDM decay are small. We again observe that as we increase the DWDM mass, the prediction of the DWDM model approaches those of DCDM.
We therefore expect cosmological constraints on DWDM to interpolate between the DCDM and $\Neff$ models.

\begin{table}
\centering
\begin{tabular}{|l|c|c|c|c|}
      \cline{2-4} 
 \hline 
               Planck Only      & $m_x=1$ eV & $m_x=10$ eV & $m_x=40$ eV  \\
\hline 
100$\theta_{s}$      & $1.04169_{-0.00034}^{+0.00034}$ & $1.04182^{+0.00036}_{-0.00036} $ & $1.04175^{+0.00034}_{-0.00034}$ \\
100$\omega_{b}$      & $2.245^{+0.015}_{-0.018}$ & $2.249^{+0.015}_{-0.017}$ & $2.251^{+0.017}_{-0.017}$ \\
$\omega_{cdm}$       &  $0.1214^{+0.0014}_{-0.0019}$ & $0.1215^{+0.0014}_{-0.0018}$   &$0.1219^{+0.0014}_{-0.0019}$ \\
$\ln 10^{10}A_{s}$     & $3.050_{-0.015}^{+0.015}$ & $3.050^{+0.015}_{-0.015}$ & $3.050^{+0.016}_{-0.016}$  \\
$n_{s}$              & $0.9699^{+0.0046}_{-0.0055}$ & $0.9714^{+0.0047}_{-0.0058}$     & $0.9712^{+0.0047}_{-0.0056}$\\
$\tau_{reio}$        & $0.0551_{-0.0076}^{+0.0076}$ & $0.0549^{+0.0077}_{-0.0077}$     & $0.0548^{+0.0079}_{-0.0079}$ \\
$\Neffx$     & $<0.26$ & $<0.20$ & $<0.14$ \\
$\log_{10}(\tau_{x}$/yr) & $-$ & $<4.0$  & $<3.1$ \\
\hline
$H_{0}$ [km/s/Mpc]              & $68.75^{+0.61}_{-0.90}$ & $68.77^{+0.63}_{-0.87}$   & $69.05^{+0.66}_{-0.95}$ \\
$\sigma_8$           & $0.8279^{+0.0073}_{-0.0073}$ & $0.8301^{+0.0067}_{-0.0080}$ & $0.8301^{+0.0079}_{-0.0079}$  \\\hline
$\chi_{min}^2$         & 1011.87  & 1011.90    & 1011.73 \\ \hline 
\end{tabular}
\caption{Constraints on the cosmological parameters from Planck~\cite{Aghanim:2019ame} data, including the lifetime, $\tau_x$, and initial abundance, $\Neffx$, of the decaying warm dark matter (DWDM). Here, the quantity $\Neffx$ reflects the energy density of DWDM at early times ($t \ll \tau_{x}$ and while DWDM is still relativistic). The uncertainties on the means shown are at the 1$\sigma$ level, upper limits are 2$\sigma$, and the $\chi^2$ values correspond to the maximum of the total likelihood. The posterior distribution for $\log_{10} (\tau_x/\mathrm{yr})$ is broad in the $m_x = 1\;\eV$ case so we do not give a summary statistic. \label{tab:planckconstraints}}
\end{table}

\begin{table}
\centering
\begin{tabular}{|l|c|c|c|c|}
      \cline{2-4} 
 \hline 
 Planck+BAO      & $m_x=1$ eV & $m_x=10$ eV & $m_x=40$ eV  \\
\hline 
100$\theta_{s}$    & $1.04170^{+0.00033}_{-0.00033}$ & $1.04184^{+0.00036}_{-0.00036}$ & $1.04174^{+0.00034}_{-0.00033}$   \\
100$\omega_{b}$      & $2.245^{+0.014}_{-0.015}$ & $2.250^{+0.015}_{-0.015}$  & $2.249^{+0.015}_{-0.015}$ \\
$\omega_{cdm}$ & $0.1211^{+0.0011}_{-0.0017}$ & $0.1215^{+0.0012}_{-0.0017}$ & $0.1221^{+0.0013}_{-0.0020}$ \\
$\ln 10^{10}A_{s}$    & $3.050^{+0.013}_{-0.015}$ & $3.051^{+0.015}_{-0.015}$ & $3.050^{+0.014}_{-0.015}$  \\
$n_{s}$          & $0.9694^{+0.0040}_{-0.0047}$ & $0.9722^{+0.0045}_{-0.0057}$ & $0.9704^{+0.0042}_{-0.0050}$  \\
$\tau_{reio}$       & $0.0553^{+0.0067}_{-0.0074}$ & $0.0552^{+0.0073}_{-0.0073}$ & $0.0547^{+0.007}_{-0.0074}$  \\
$\Neffx$     & $<0.24$ & $<0.22$ & $<0.15$ \\
$\log_{10}(\tau_{x}$/yr) & $-$ & $< 4.1$ & $< 3.1$ \\
\hline
$H_{0}$ [km/s/Mpc]  & $68.70^{+0.46}_{-0.70}$  & $68.86^{+0.46}_{-0.69}$ & $70.00^{+0.73}_{-0.89}$ \\
$\sigma_8$           & $0.8273^{+0.0065}_{-0.0073}$  & $0.8306^{+0.0068}_{-0.0083}$ &  $0.8302^{+0.0070}_{-0.0082}$   \\  \hline 
$\chi_{min}^2$       & 1018.98 & 1019.30 &  1019.11 \\ \hline 
\end{tabular}
\caption{Constraints on the cosmological parameters from Planck~\cite{Aghanim:2019ame} combined with BAO~\cite{Alam:2016hwk,Beutler:2011hx,Ross:2014qpa} data, including the lifetime, $\tau_x$, and initial abundance, $\Neffx$, of the decaying warm dark matter (DWDM). Here, the quantity $\Neffx$ reflects the energy density of DWDM at early times ($t \ll \tau_{x}$ and while DWDM is still relativistic). The uncertainties on the means shown are at the 1$\sigma$ level, upper limits are 2$\sigma$, and the $\chi^2$ values correspond to the maximum of the total likelihood. The posterior distribution for $\log_{10} (\tau_x/\mathrm{yr})$ is broad in the $m_x = 1\;\eV$ case so we do not give a summary statistic -- see Fig.~\ref{fig:neff_vs_tau_Planck_BAO} for an illustration.\label{tab:baoconstraints}}
\end{table}

\begin{table}
\centering
\begin{tabular}{|l|c|c|c|c|}
      \cline{2-4}
 \hline 
Planck + BAO + $H_0$                      & $m_x=1$ eV & $m_x=10$ eV & $m_x=40$ eV  \\
\hline 
100$\theta_{s}$      & $1.04144^{+0.00045}_{-0.00039}$ & $1.0424^{+0.00044}_{-0.00047}$ & $1.0416^{+0.00041}_{-0.00039}$  \\
100$\omega_{b}$      & $2.262^{+0.017}_{-0.017}$ & $2.264^{+0.015}_{-0.016}$ & $2.265^{+0.015}_{-0.015}$ \\
$\omega_{cdm}$       & $0.1230^{+0.0022}_{-0.0029}$ & $0.123^{+0.0020}_{-0.0026}$ & $0.124^{+0.0020}_{-0.0029}$  \\
$\ln 10^{10}A_{s}$     & $3.059^{+0.015}_{-0.015}$ & $3.059^{+0.015}_{-0.016}$ & $3.057^{+0.0150}_{-0.0154}$  \\
$n_{s}$              & $0.9762^{+0.0055}_{-0.0062}$ & $0.977^{+0.0052}_{-0.0064}$ & $0.976^{+0.0049}_{-0.0054}$  \\
$\tau_{reio}$        & $0.0574^{+0.0073}_{-0.0073}$ & $0.0574^{+0.0072}_{-0.0079}$ & $0.0569^{+0.0076}_{-0.0070}$  \\
$\Neffx$     & $< 0.48$ & $<0.39$ & $<0.27$  \\
$\log_{10}(\tau_{x}$/yr)& $-$ & $< 3.8$ & $< 3.0$ \\
\hline
$H_{0}$ [km/s/Mpc]              & $70.04^{+0.75}_{-0.99}$ & $69.96^{+0.76}_{-0.89}$ & $70.20^{+0.79}_{-0.94}$ \\
$\sigma_8$           & $0.8331^{+0.0094}_{-0.0094}$ & $0.8354^{+0.0083}_{-0.0095}$ & $0.836^{+0.0086}_{-0.0093}$    \\ \hline 
$\chi_{min}^2$       & 1031.47 & 1031.86 & 1031.90  \\ \hline 
\end{tabular}
\caption{Constraints on the cosmological parameters from Planck~\cite{Aghanim:2019ame}, BAO~\cite{Alam:2016hwk,Beutler:2011hx,Ross:2014qpa}, and local $H_0$~\cite{Riess:2019cxk} data, including the lifetime, $\tau_x$, and initial abundance, $\Neffx$, of the decaying warm dark matter (DWDM). Here, the quantity $\Neffx$ reflects the energy density of DWDM at early times ($t \ll \tau_{x}$ and while DWDM is still relativistic). The uncertainties on the means shown are at the 1$\sigma$ level, upper limits are 2$\sigma$, and the $\chi^2$ values correspond to the maximum of the total likelihood. Note that even though we quote upper limits on $\Neffx$, the posterior distributions for this quantity display a $\sim 1\sigma$ preference for non-zero values. The posterior distribution for $\log_{10} (\tau_x/\mathrm{yr})$ is broad in the $m_x = 1\;\eV$ case so we do not give a summary statistic.\label{tab:h0andbaoconstraints}}
\end{table}

\section{Results and Discussion}
\label{sec:results}

In this section, we implement the 
model described in Sec.~\ref{sec:boltzmann} in the standard \CLASS and \MP~\cite{Audren:2012wb,Brinckmann:2018cvx} (version 3.2)
pipeline in order to establish which regions of parameter space are empirically viable. We employ the following data sets to constrain this parameter space:
\begin{itemize}
    \item The 2018 Planck measurements of the CMB (via \texttt{TTTEEE} \texttt{Plik lite} high-$\ell$, \texttt{TT} and \texttt{EE} low-$\ell$, and lensing likelihoods)~\cite{Aghanim:2019ame},
    \item BAO data from the BOSS survey (data release 12)~\cite{Alam:2016hwk}, low-redshift measurements from the 6dF survey~\cite{Beutler:2011hx}, and the BOSS main galaxy sample~\cite{Ross:2014qpa},
    \item The local measurement of the Hubble constant, $H_0=74.03\pm 1.42\;\mathrm{km/s/Mpc}$~\cite{Riess:2019cxk}.
\end{itemize}
Our results are obtained by running 8 chains for each model and monitoring convergence until the Gelman-Rubin~\cite{Gelman:1992zz} criterion, $R -1 < 0.05$, is satisfied for all of the parameters. In addition to the standard cosmological parameters $\{\theta_s, \omega_b, \omega_{cdm}, \ln 10^{10} A_s, n_s, \tau_{reio}\}$~\cite{Aghanim:2018eyx}, 
we scan over the DWDM initial abundance $\Neffx$ and lifetime $\log_{10}(\tau_x/\mathrm{yr})$ (with the latter having a flat prior in the range $[2,6]$). We consider three different masses for the DWDM particle (taken to be a Weyl fermion), $m_x = 1,\;10$ and $40$ eV, with the initial DWDM temperature determined self-consistently from the initial abundance via Eq.~\ref{eq:dwdm_initial_abundance_from_equilibrium} (an alternative possibility is 
  to fix $T_x = T_\nu$ and treat $\Neffx$ as an independent parameter as discussed in Sec.~\ref{sec:model} -- this case is qualitatively similar 
  to our choice since the fourth root relating $T_x$ and $\Neffx$ in Eq.~\ref{eq:dwdm_initial_abundance_from_equilibrium} ensures that $T_x \sim T_\nu$ in most of the 
parameter space anyway). These three masses span the range of free-streaming scales that are relevant for the observed CMB, with the largest mass approaching that of the cold decaying dark matter regime studied in Ref.~\cite{Poulin:2016nat}. We note that for lighter masses and shorter 
lifetimes the backreaction from inverse decays can be important as discussed around Eq.~\ref{eq:nr_decay_condition}. This backreaction 
would prevent free-streaming of DWDM and the decay product DR; this regime is beyond the scope of our analysis, but this caveat should be 
kept in mind when interpreting the $m_x = 1\;\eV$ results.

\begin{figure}
    \centering
    \includegraphics[width=0.47\textwidth]{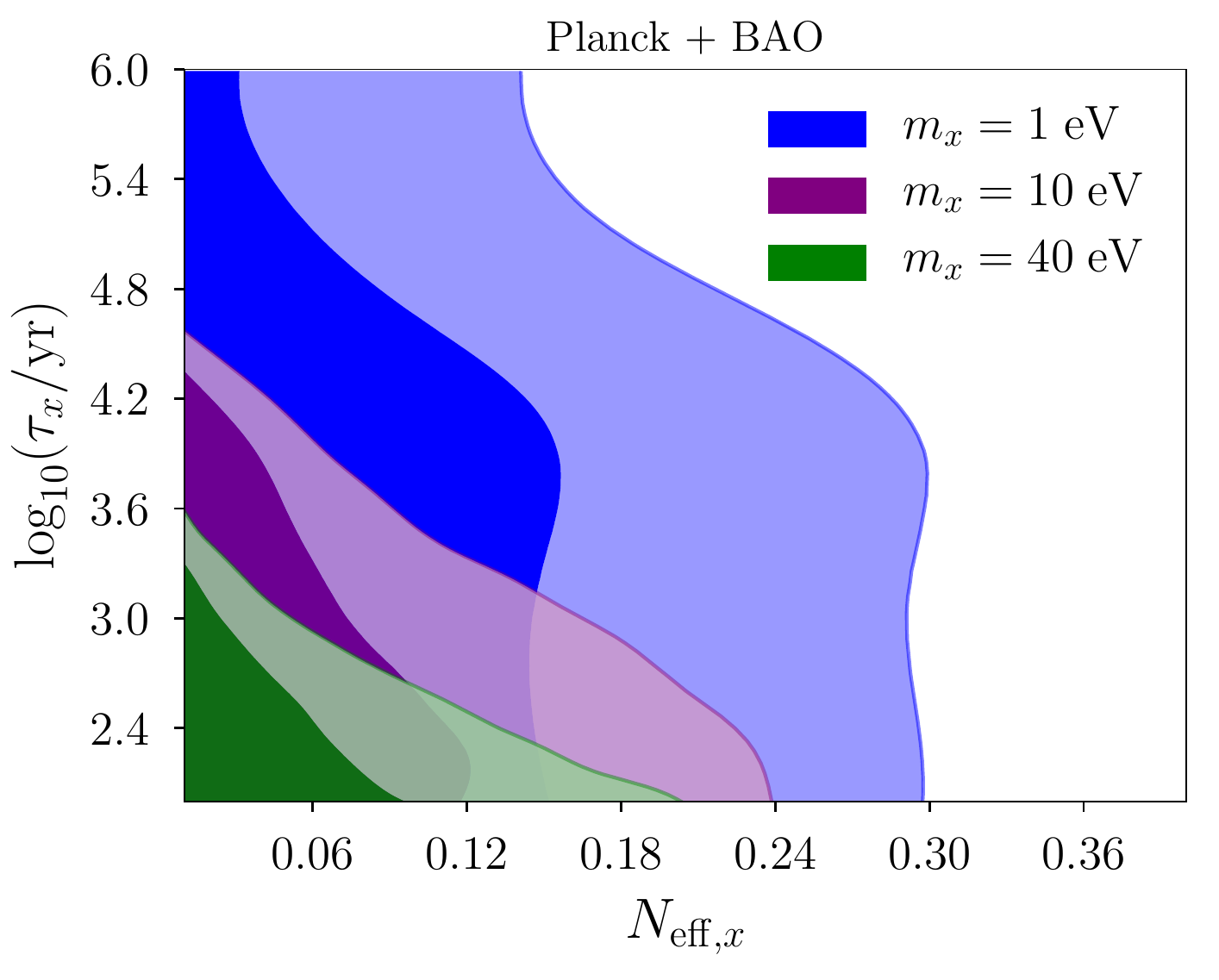}
    \caption{The regions of parameter space that are favored by the Planck and BAO data in models that include a component of decaying warm dark matter (DWDM), with a mass of $m_x=1$, $10$ or $40$ eV. Here, $\tau_x$ denotes the lifetime of the DWDM, while $\Neffx$ reflects the energy density of DWDM at early times ($t \ll \tau_{x}$ and while DWDM is still relativistic). The light (dark) colored regions represent the 1$\sigma$ (2$\sigma$) confidence regions.}
    \label{fig:neff_vs_tau_Planck_BAO}
\end{figure}

In a series of tables, we show the constraints on the cosmological parameters (including the DWDM abundance and lifetime), utilizing the Planck data (Table~\ref{tab:planckconstraints}), the Planck and BAO data (Table~\ref{tab:baoconstraints}), and the Planck, BAO and local $H_0$ data (Table~\ref{tab:h0andbaoconstraints}). In Appendix.~\ref{appendixA}, we present a similar set of tables, containing results for \lcdm (Table~\ref{tab:lcdm_results}), and an extension of \lcdm with simple dark radiation parametrized by $\Neff$ (Table~\ref{tab:neff_results}). From these tables, we see that Planck and BAO are consistent with a modest abundance of DWDM but do not exhibit a strong preference for any particular mass or lifetime. 
This is illustrated in Fig.~\ref{fig:neff_vs_tau_Planck_BAO}, where 
we show the posterior distribution in the $\Neffx - \tau_x$ plane for the Planck + BAO data set, and for three different choices of the DWDM masses, $m_x$. We note that 
larger DWDM masses are constrained to have shorter lifetimes and smaller abundances. 
This is the natural expectation since a larger mass implies an earlier non-relativistic transition and a longer period of growth for DWDM density relative to that of radiation, resulting in a larger injection of dark radiation. A similar 
argument holds for longer lifetimes.

Tables~\ref{tab:planckconstraints}, \ref{tab:baoconstraints}, \ref{tab:h0andbaoconstraints}, \ref{tab:lcdm_results} and~\ref{tab:neff_results} also present the overall quality of these fits, as measured by 
the effective minimum $\chi^2$. The quality of these fits to the Planck and BAO data are similar, regardless of whether or not a component of DWDM is included. This is not surprising, since these data sets do not show any significant preference for an additional component of energy.
This changes significantly when we include the local measurement of $H_0$ in the fit, as we discuss in more detail below.   

\begin{figure}[t!]
    \centering
    \includegraphics[width=0.47\textwidth]{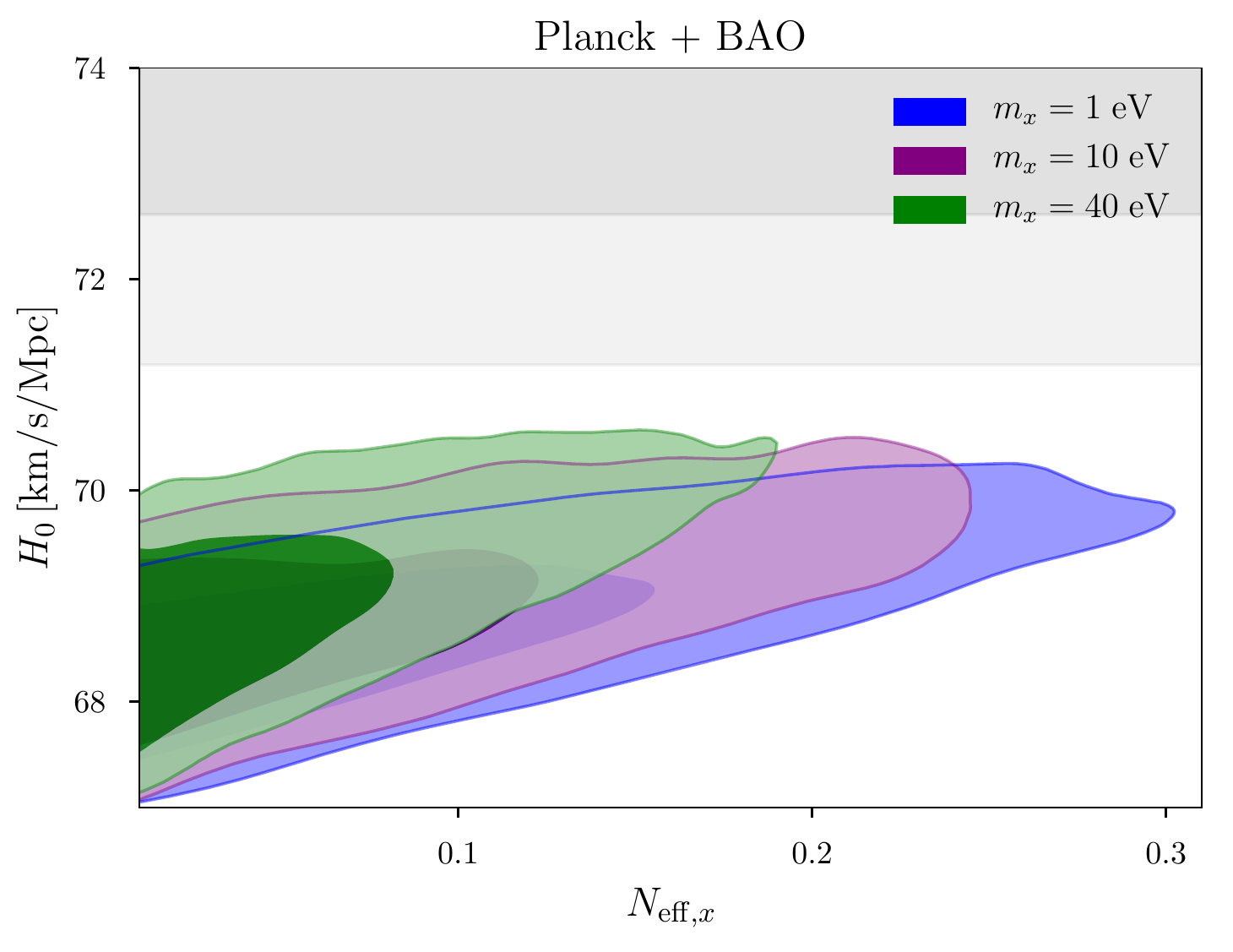}
    \includegraphics[width=0.47\textwidth]{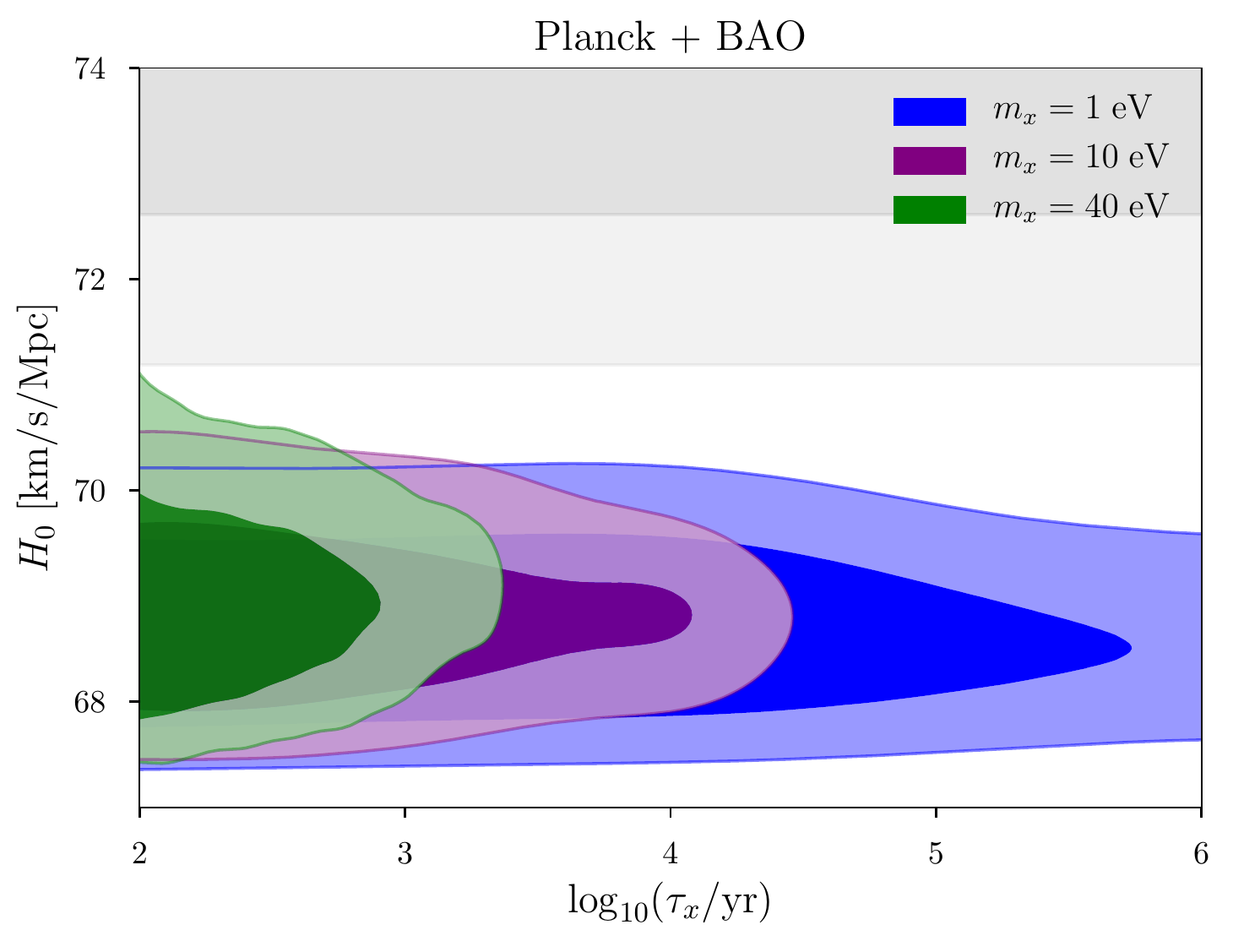}\\
    \includegraphics[width=0.47\textwidth]{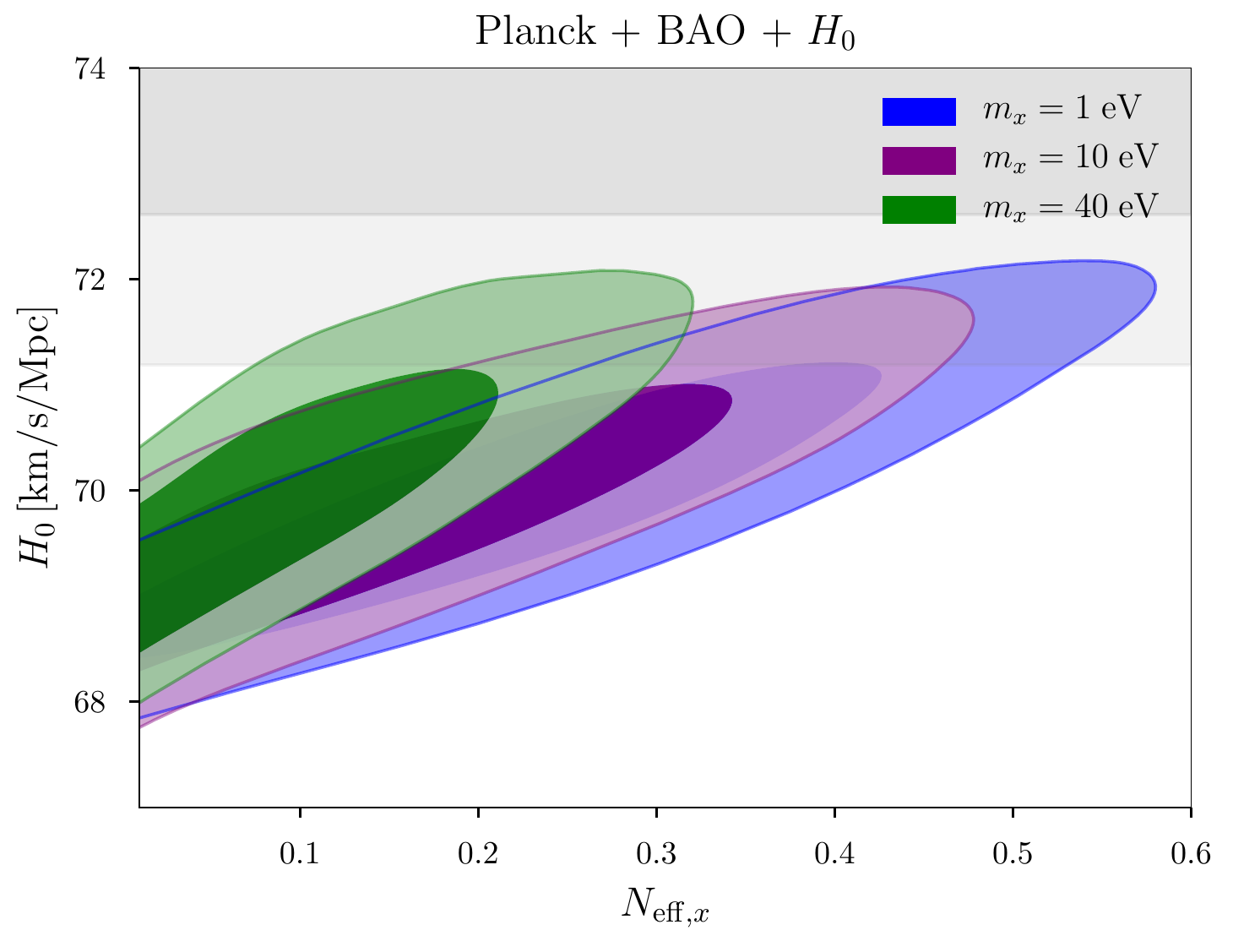}
    \includegraphics[width=0.47\textwidth]{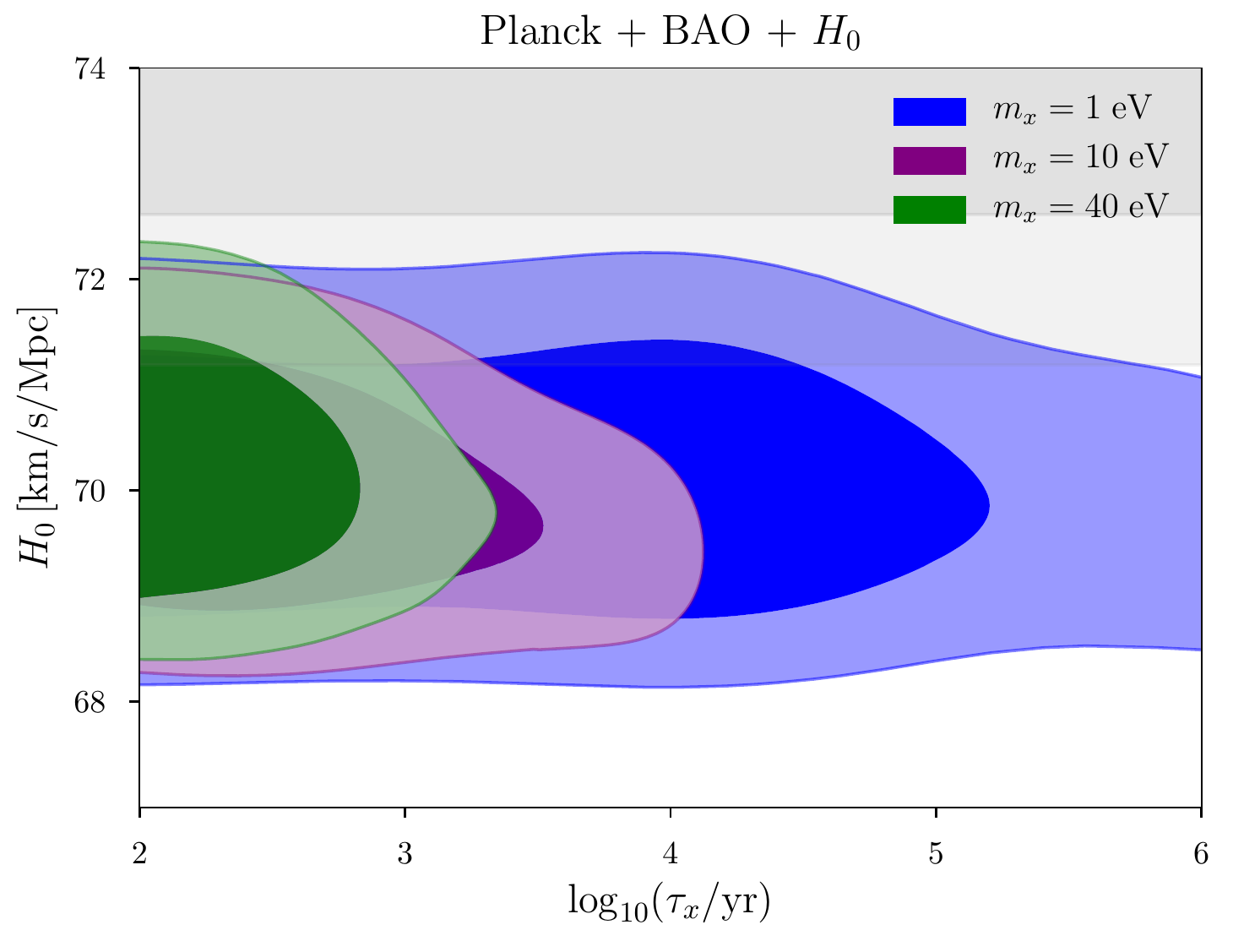}
    \caption{The regions of parameter space that are favored by the data in models that include a component of decaying warm dark matter (DWDM), with a mass of $m_x=1$, $10$ or $40$ eV. Here, $\tau_x$ denotes the lifetime of the DWDM, while $\Neffx$ reflects the energy density of DWDM at early times ($t \ll \tau_{x}$ and while DWDM is still relativistic). In the upper panels, we include only Planck and BAO data, while in the lower panels we also incorporate a likelihood for the local $H_0$ measurement. The light (dark) colored regions represent the 1$\sigma$ (2$\sigma$) confidence regions. The light (dark) gray shaded regions denote the 68\% (95\%) confidence interval of the local $H_{0}$ measurement of Ref.~\cite{Riess:2019cxk}.}
    \label{fig:H0vsnefftauBAO}
\end{figure}

In Fig~\ref{fig:H0vsnefftauBAO}, we plot the regions of the DWDM parameter space that are compatible with two different combinations of cosmological data. We find that these datasets are consistent with a modest quantity of DWDM (corresponding to $\Neffx$ as large as $\sim 0.1-0.3$ at early times), with a lifetime as long as roughly $\sim 10^3-10^6$ years (depending on the value of $m_x$).
In Appendix~\ref{sec:triangle_plots}, we present the full set of 2D posterior distributions, which illustrate the correlations between the various parameters in greater detail.

Note that, within the context of standard $\Lambda$CDM cosmology, the combination of Planck and BAO data require $H_0 < 68.9$ km/s/Mpc at 2$\sigma$ confidence. In contrast, the 2$\sigma$ contours in the upper panels of Fig.~\ref{fig:H0vsnefftauBAO} extend up to $\sim 70-71$ km/s/Mpc, significantly closer to the regions favored by local $H_0$ measurements~\cite{Riess:2019cxk} (shown as grey horizontal bands). 
In the lower panels of Fig.~\ref{fig:H0vsnefftauBAO} we combine the Planck + BAO results with the local measurement of $H_0$. We see that the posterior distributions extend to much larger values of $H_0$ compared to \lcdm for the same 
data combination (see Table~\ref{tab:lcdm_results}). We note that larger values of $\Neffx$ are associated with 
larger $H_0$ as expected from the impact of extra energy density on the 
sound horizon~\cite{Aylor:2018drw,Knox:2019rjx}.
This illustrates the potential of DWDM to relieve the tension between local and cosmological determinations of the Hubble constant.

\begin{figure}
    \centering
    \includegraphics[width=0.47\textwidth]{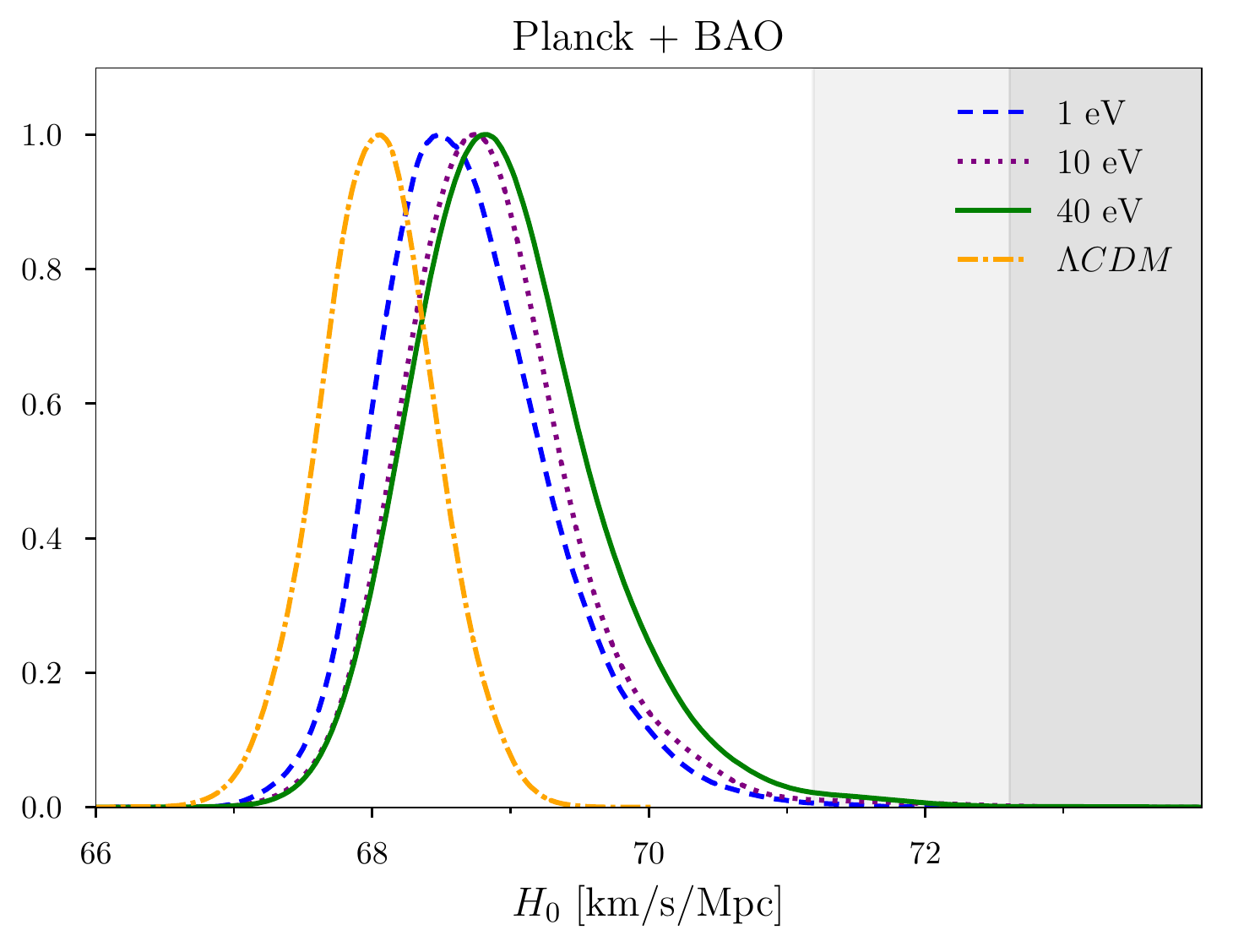}
    \includegraphics[width=0.47\textwidth]{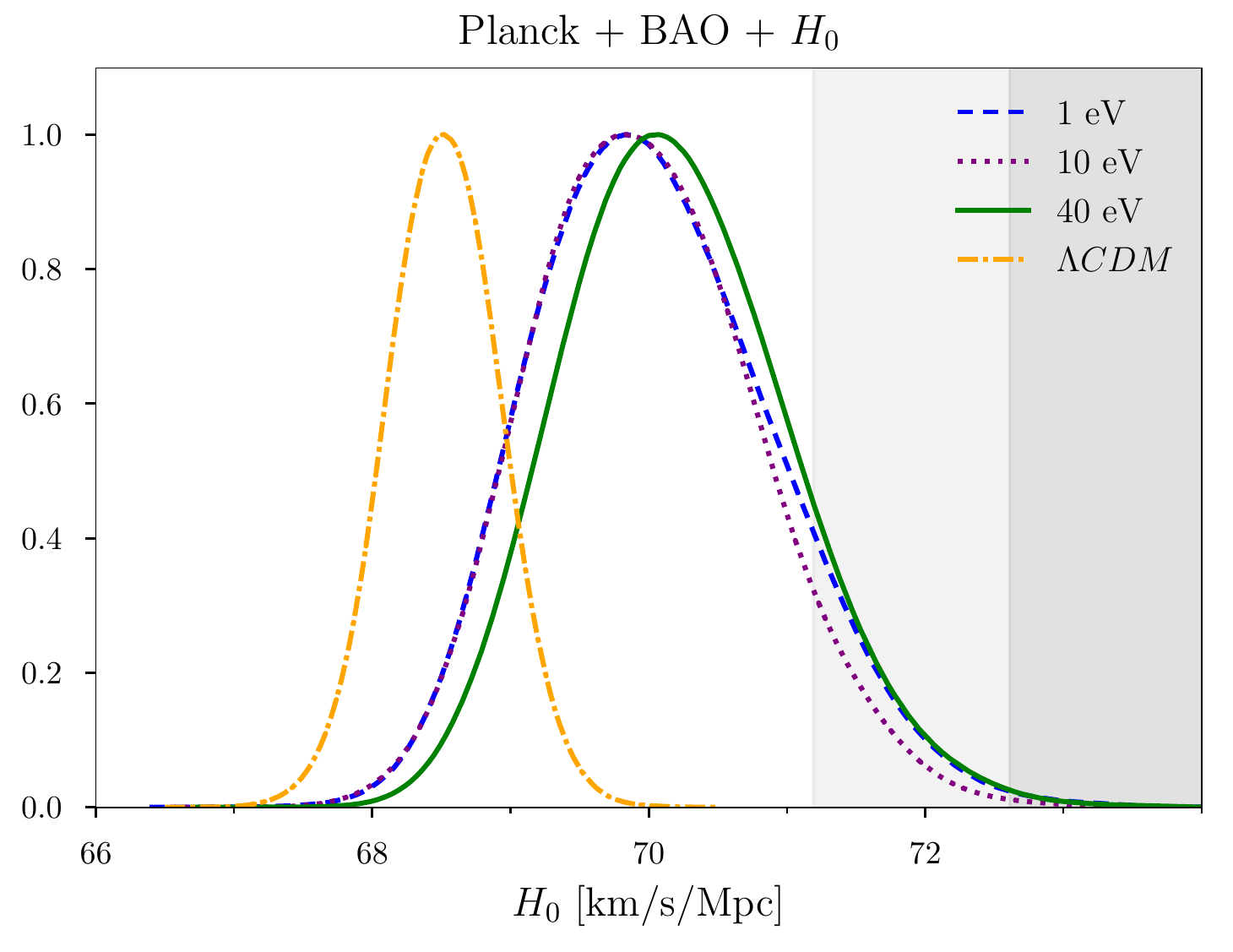}
    \caption{The marginalized posterior distribution of $H_{0}$ in \lcdm and in a model with a component of decaying warm dark matter (DWDM) with a mass of $m_x=1$, $10$ or $40$ eV. In the left panel, we use only Planck and BAO data, while in the right we also include a likelihood 
    for the local $H_0$ measurement. The light (dark) gray shaded regions indicate the 68\% (95\%) confidence interval of the local $H_{0}$ measurement of Ref.~\cite{Riess:2019cxk}.}
    \label{fig:H0_posterior}
\end{figure}

This conclusion is more directly demonstrated in Fig.~\ref{fig:H0_posterior}, where we show the main result of our analysis. In particular, we plot the marginalized posterior distribution of $H_0$ in \lcdm, and in a model that includes a component of DWDM (for three choices of $m_x$). This figure makes it clear that the presence of a DWDM component shifts the range of $H_0$ values that are favored by Planck+BAO data upward and toward those preferred by local measurements. Furthermore, the tails of this distribution become broader when a component of DWDM is included, and thus overlap to a larger degree with the range favored by local measurements. More quantitatively, we find that the tension between the Planck+BAO data and the local Hubble measurement is 4.03$\sigma$ in the \lcdm framework. When a component of DWDM is included, this tension is reduced to 2.9$\sigma$ (for each of $m_x=1$ eV, 10 eV and 40 eV). A second way to see the improved fit obtained in the DWDM scenario is 
to compare the minimum value of the effective $\chi^2$ for the Planck+BAO+$H_0$ 
data combination in Tables~\ref{tab:baoconstraints},~\ref{tab:lcdm_results} and~\ref{tab:neff_results}. We obtain a significant decrease in $\chi^2$ of $3.1-3.7$ compared to \lcdm, which is comparable to the $\Neff$ model (albeit with one more model parameter).

\section{Discussion and Conclusions}
\label{sec:conclusion}

The disagreement between local measurements of the Hubble constant and the value of this quantity inferred from the CMB and other cosmological probes has provided motivation for a variety of extensions of the standard $\Lambda$CDM model, most of which involve the presence of additional energy at or around the time of matter-radiation equality. In this study, we have considered the possibility that there may have existed a modest component of semi-relativistic dark matter that decayed into dark radiation during this era. This is representative of a broad class of scenarios in which the expansion rate is increased (relative to that predicted in \lcdm) during the period prior to recombination, leading to a reduction in the sound horizon and to a larger inferred value of $H_0$.

Scenarios that feature a component of decaying warm dark matter are easily realizable within the context of simple and well-motivated particle physics models. Unstable sterile neutrinos that decay into particles within a dark sector are one particularly attractive possibility. Furthermore, decaying warm dark matter particles can have a qualitatively different impact on the evolution of perturbations than is predicted in the case of decaying cold dark matter. Before these unstable particles become non-relativistic, they act as dark radiation. When they decay, they transfer their energy into dark radiation once again. There is thus a finite window (depending on the mass and lifetime) during which the combination of decaying warm dark matter and dark radiation evolves non-trivially. We find that this non-trivial evolution of the equation of state and the corresponding impact on cosmological perturbations can result in the relaxation of the $H_0$ tension, from over $4\sigma$ in \lcdm to approximately $2.9\sigma$. This is similar to what can be accomplished by introducing an additional component of dark radiation.

Independent of the Hubble tension, this work provides a useful framework for studying decaying, semi-relativistic relics in the early universe. In particular, our analysis enables the application of observational probes that are sensitive to the evolution of cosmological perturbations. As a result, the constraints presented here are superior to background-only bounds on such relics (for example, those derived from the light element abundances alone), and are applicable to a wide range of masses and lifetimes. Despite the precision of modern cosmological observations, a modest component of decaying warm dark matter remains compatible with the data, motivating further studies of this class of models.

\section*{Acknowledgments}
We would like to thank Gustavo Marques-Tavares, Patrick Draper, Jonathan Kozaczuk, Albert Stebbins, Gordan Krnjaic, Ken Van Tilburg, Julia Stadler,
Massimiliano Lattanzi, Vivian Poulin, Yuhsin Tsai, Martina Gerbino and Thejs Brinckmann for useful discussions, and 
Sam McDermott for collaboration at early stages of this work.
This manuscript has been authored by Fermi Research Alliance, LLC under Contract No. DE-AC02-07CH11359 with the U.S. Department of Energy, Office of Science, Office of High Energy Physics. We thank the Galileo Galilei Institute for Theoretical Physics for the hospitality and the INFN for partial support during the completion of this work. This material is based upon work supported by the National Science Foundation Graduate Research Fellowship Program under Grant No. DGE-1746045. Any opinions, findings, and conclusions or recommendations expressed in this material are those of the author(s) and do not necessarily reflect the views of the National Science Foundation. 

\appendix
\section{Dark Radiation Source Terms}
\label{sec:dr_source_terms}
In this section we provide details of the calculation of the DR collision term that 
involve integrals $I$ and $I_F$ defined in Eqs.~\ref{eq:dr_collision_int0} and~\ref{eq:dr_collision_int1}, respectively.
In order to simplify $I(p_1)$, we will use the spatial part of the $\delta$-function to perform the $d\Pi_2$ integral, 
and then rewrite the remaining energy in terms of $p_1 = E_1$:
\begin{align}
  I(p_1) & = \frac{1}{2E_1} \int d\Pi_3 \left(\frac{2\pi}{2E_2}\right) \delta[E_3 - E_1 - E_2] f(\vec{k},q_3,\hat{n}_3,\tau) \\
  & = \frac{1}{2E_1}\int d\Pi_3 \left(\frac{2\pi}{2E_2}\right) \left(\frac{2 E_1 E_2}{m^2_x}\right)\delta\left[E_1 - \frac{m^2_x}{2(E_3 - p_3 c_{13})}\right] f \\
  & = \frac{\pi}{m^2_x}\int d\Pi_3 \delta\left[E_1 - \frac{m^2_x}{2(E_3 - p_3 c_{13})}\right] f(p_3), 
\end{align}
where $f$ is the distribution of the decaying particle (in Sec.~\ref{sec:DWDM_perturb}, $q_3$ and $\hat{n}_3$ were just called $q$ and $\hat{n}$) and we used $E_2 = \sqrt{p_3^2 + p_1^2 - 2 p_1 p_3 c_{13}}$, where 
$c_{13} = \cos\theta_{13}$ is the angle between $p_1$ and $p_3$.

We can now evaluate $I_F$, defined in Eq.~\ref{eq:dr_collision_int1}, by using the remaining $\delta$ function to carry out the $p_1 = E_1$ integration:
\beq
  I_F = \frac{\pi m^4_x}{8} \int d\Pi_3 \frac{f^{(0)}(p_3)(1 + \Psi)}{(E_3 - p_3 c_{13})^3} \equiv I_F^{(0)} + I_F^{(1)},
\eeq
where in the last step we separated the background and perturbed contributions. 
The background distribution, $f^{(0)}$, only depends on 
the magnitude of the DWDM momentum, so the angular integrals in $I_F^{(0)}$ can be carried out explicitly:
\begin{align}
  I_F^{(0)} & = \frac{\pi m^4_x}{8} \int \frac{dp_3 p_3^2}{(2\pi)^3 2E_3} f^{(0)}(p_3) \int d\phi dc_{13} \frac{1}{(E_3 - p_3 c_{13})^3} \\
  & = \frac{1}{32\pi} \int dp_3 p_3^2 f^{(0)}(p_3) \nonumber \\
& = \frac{\pi}{16} n,
\end{align}
which is the naive expectation as discussed in Sec.~\ref{sec:dr_perturb}.

In order to simplify $I_F^{(1)}$, we make use of the Legendre decomposition of the DWDM perturbation, $\Psi$, given in Eq.~\ref{eq:DWDM_legendre_decomp}.
The angular dependence is restricted to the Legendre polynomials (and only through $\hat k \cdot \hat{n}_3$), whereas the natural 
integration variables above are $\phi$ and $c_{13}$. To proceed, we pick a coordinate system such that 
\begin{align}
  \hat{n}_1 & = (0,0,1) \\
  \hat{n}_3 & = (\sin\theta_{13}\cos\phi,\sin\theta_{13}\sin\phi,\cos\theta_{13})  \\
  \hat{k} & = (\sin\theta_1, 0, \cos\theta_1). 
\end{align}
Then, the relevant dot products are 
\begin{align}
  \hat k \cdot \hat n_1 & = \cos\theta_1 \\
  \hat k \cdot \hat n_3 & = \cos\theta_3 = \sin\theta_1\sin\theta_{13}\cos\phi + \cos\theta_1\cos\theta_{13}\\
  \hat n_1\cdot \hat n_3 & = \cos\theta_{13}.
\end{align}
Note that $\phi$-dependence enters only through $\hat k\cdot \hat n_3$, which appears in the Legendre decomposition 
of $\Psi$. We can therefore use the following identity~\cite{NIST:DLMF}:
\beq
\int_0^{2\pi} d\phi P_l(\sin\theta_1\sin\theta_{13}\cos\phi + \cos\theta_1\cos\theta_{13}) = 2\pi P_l (\cos\theta_1) P_l(\cos\theta_{13})
\eeq
to do the $\phi$ integral. This yields:
\begin{align}
  \label{eq:dr_collision_int2}
  I_F^{(1)} & = \sum_{l=0}^{\infty} (-i)^l (2l+1) P_l(\hat k\cdot \hat n_1) \times \frac{\pi m^4_x}{8} \int \frac{dp_3 p_3^2}{(2\pi)^2 2E_3} f^{(0)}\Psi_l \int dc_{13} \frac{P_l(c_{13})}{(E_3 - p_3 c_{13})^3}. 
\end{align}
The $dc_{13}$ integral can be performed for any $l$ analytically, but it is useful to first rewrite it as follows: 
\begin{align}
  \int dc_{13} \frac{P_l(c_{13})}{(E_3 - p_3 c_{13})^3} & = \frac{1}{E_3^3} \int dc_{13} \frac{P_l(c_{13})}{(1 - (p_3/E_3) c_{13})^3} \\ 
  & \equiv \frac{2}{E_3^3(1 - p_3^2/E_3^2)^2}  \mathcal{F}_l(p_3/E_3) \\ 
  & = \frac{2E_3}{m^4_x}  \mathcal{F}_l(p_3/E_3),
\end{align}
where in the last two lines we defined
\beq
\mathcal{F}_l(x) = \frac{(1-x^2)^2}{2}\int_{-1}^{+1} \frac{du P_l(u)}{(1 - xu)^3}.
\eeq
The normalization is such that $\mathcal{F}_0(x) = 1$ and $\mathcal{F}_{1}(x) = x$. These functions have the following useful limiting forms:
\begin{align}
  \mathcal{F}_l(x) & \sim \frac{2^{l-1} (l!)^2 (l+1)(l+2)}{(2l+1)!} x^l \;\;\; (x\ll 1) \label{eq:F_limiting_form}\\
\mathcal{F}_l(x) & \sim 1 + \mathcal{O}(x-1) \;\;\; (|x-1|\ll 1). 
\end{align}
Plugging the above definition into Eq.~\ref{eq:dr_collision_int2} yields Eq.~\ref{eq:dr_collision_int3}.

\section{Parameter Estimates for \texorpdfstring{$\Lambda$}{Lambda}CDM and \texorpdfstring{$\Neff$}{Neff} Models}
\label{appendixA}

In this Appendix, we present the results of our Monte Carlo for the case of \lcdm, and for \lcdm with an additional component of dark radiation (i.e. $\Neff$). 
We follow the same procedure as outlined in Sec.~\ref{sec:results}. 
The parameter means and their uncertainties, along with the best-fit values of $\chi^2$ are shown in Table~\ref{tab:lcdm_results} for \lcdm, and in Table~\ref{tab:neff_results} for $\Neff$.

\begin{table}
  \centering
  \begin{tabular}{|l|c|c|c|}
\hline
 &  Planck  & Planck+BAO  &  Planck+BAO+$H_0$ \\
\hline
$100\theta_s$  &  $1.0419^{+0.00029}_{-0.00031}$  &  $1.0419^{+0.00028}_{-0.00029}$  &  $1.042^{+0.00028}_{-0.00029}$\\
$100\omega_b$  &  $2.237^{+0.015}_{-0.015}$   &  $2.239^{+0.014}_{-0.013}$       &  $2.249^{+0.014}_{-0.014}$\\
$\omega_{cdm}$    &  $0.1199^{+0.0012}_{-0.0013}$  &  $0.11961^{+0.00093}_{-0.00095}$  &  $0.11863^{+0.00092}_{-0.00093}$\\
$\ln 10^{10}A_s$  &  $3.043^{+0.014}_{-0.015}$    &  $3.044^{+0.014}_{-0.014}$       &  $3.049^{+0.014}_{-0.015}$\\
$n_s$           &  $0.965^{+0.0041}_{-0.0044}$    &  $0.9656^{+0.0038}_{-0.0036}$    &  $0.9681^{+0.0037}_{-0.0037}$\\
$\tau_{reio}$         &  $0.0539^{+0.0073}_{-0.0078}$    &  $0.0546^{+0.0071}_{-0.0073}$    &  $0.0576^{+0.0069}_{-0.0075}$\\
\hline
$H_0$ [km/s/Mpc]              &  $67.9^{+0.58}_{-0.55}$     &  $68.04^{+0.42}_{-0.43}$         &  $68.52^{+0.41}_{-0.42}$\\
$\sigma_8$          &  $0.8232^{+0.0059}_{-0.0062}$  &  $0.8227^{+0.0057}_{-0.0063}$  &  $0.8217^{+0.006}_{-0.0062}$\\
\hline
$\chi^2_{min}$ & 1011.61  &  1018.79  &  1035.19 \\
\hline
  \end{tabular}
  \caption{Constraints on the cosmological parameters in the standard \lcdm model from Planck data (left), Planck and BAO data (center), or Planck, BAO and local $H_0$ data (right). The uncertainties on the mean values are given at the 1$\sigma$ level, and the data sets employed are described in Sec.~\ref{sec:results}. The $\chi^2$ values shown correspond to the maximum of the total likelihood for all data sets.\label{tab:lcdm_results}}
\end{table}

\begin{table}
  \centering
  \begin{tabular}{|l|c|c|c|}
\hline
 &  Planck  & Planck+BAO  &  Planck+BAO+$H_0$ \\
\hline
$100\theta_s$  &  $1.0422^{+0.00051}_{-0.00053}$  &  $1.0422^{+0.00052}_{-0.0005}$  &  $1.0414^{+0.00043}_{-0.00044}$\\
$100\omega_b$  &  $2.222^{+0.022}_{-0.022}$   &  $2.23^{+0.018}_{-0.019}$       &  $2.264^{+0.016}_{-0.016}$\\
$\omega_{cdm}$    &  $0.1176^{+0.0029}_{-0.0029}$  &  $0.1179^{+0.0028}_{-0.003}$   &  $0.123^{+0.0027}_{-0.0026}$\\
$\ln 10^{10}A_s$  &  $3.035^{+0.017}_{-0.018}$    &  $3.039^{+0.016}_{-0.016}$      &  $3.057^{+0.015}_{-0.016}$\\
$n_s$           &  $0.9586^{+0.0084}_{-0.0086}$   &  $0.962^{+0.0069}_{-0.0071}$    &  $0.9761^{+0.0058}_{-0.0056}$\\
$\tau_{reio}$         &  $0.0531^{+0.0074}_{-0.0077}$    &  $0.0543^{+0.0068}_{-0.0073}$   &  $0.057^{+0.0068}_{-0.0079}$\\
$\Neff$          &  $2.88^{+0.18}_{-0.19}$        &  $2.94^{+0.17}_{-0.18}$         &  $3.3^{+0.15}_{-0.15}$\\
\hline
$H_0$ [km/s/Mpc]               &  $66.8^{+1.3}_{-1.5}$       &  $67.4^{+1.1}_{-1.2}$           &  $69.96^{+0.94}_{-0.9}$\\
$\sigma_8$          &  $0.8163^{+0.0098}_{-0.0106}$  &  $0.8177^{+0.0099}_{-0.0099}$  &  $0.8334^{+0.009}_{-0.0092}$\\
\hline
$\chi^2_{min}$ & 1010.8  &  1018.45  &  1032.16 \\
\hline
  \end{tabular}
  \caption{Constraints on the cosmological parameters in a \lcdm+$\Neff$ model from Planck data (left), Planck and BAO data (center), or Planck, BAO and local $H_0$ data (right). The uncertainties on the mean values are given at the 1$\sigma$ level, and the data sets employed are described in Sec.~\ref{sec:results}. The $\chi^2$ values shown correspond to the maximum of the total likelihood for all data sets.\label{tab:neff_results}}
\end{table}

\section{Posterior Distributions for DWDM Models}
\label{sec:triangle_plots}

\begin{figure}
    \centering
    \includegraphics[width=\textwidth]{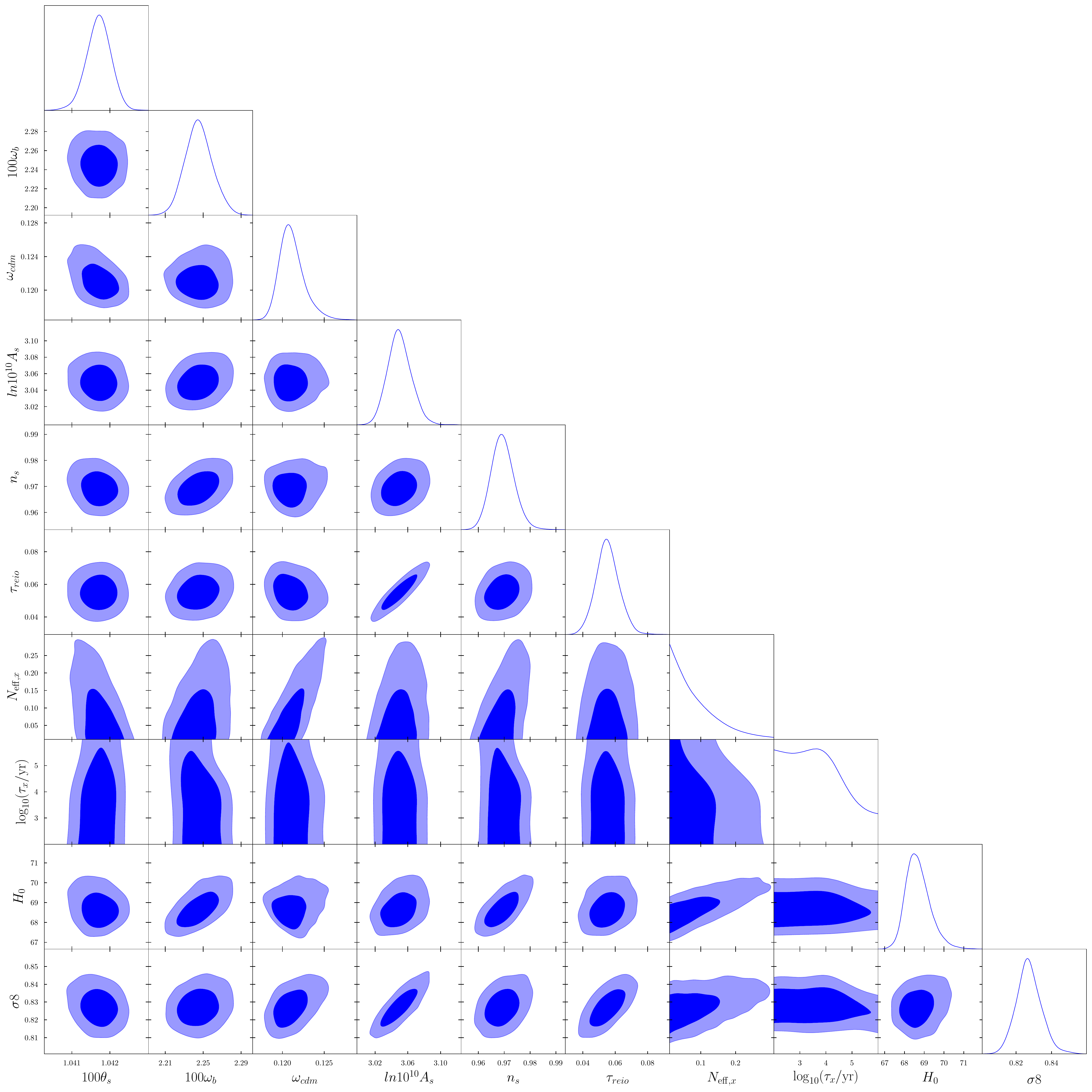}
    \caption{Posterior distributions for the DWDM model with $m_x =1$ eV, for the Planck + BAO data set. Inner darker (outer lighter) regions correspond to 
    $1\sigma$ ($2\sigma$) confidence regions.\label{fig:bao1_tri}}
\end{figure}

\begin{figure}
    \centering
    \includegraphics[width=\textwidth]{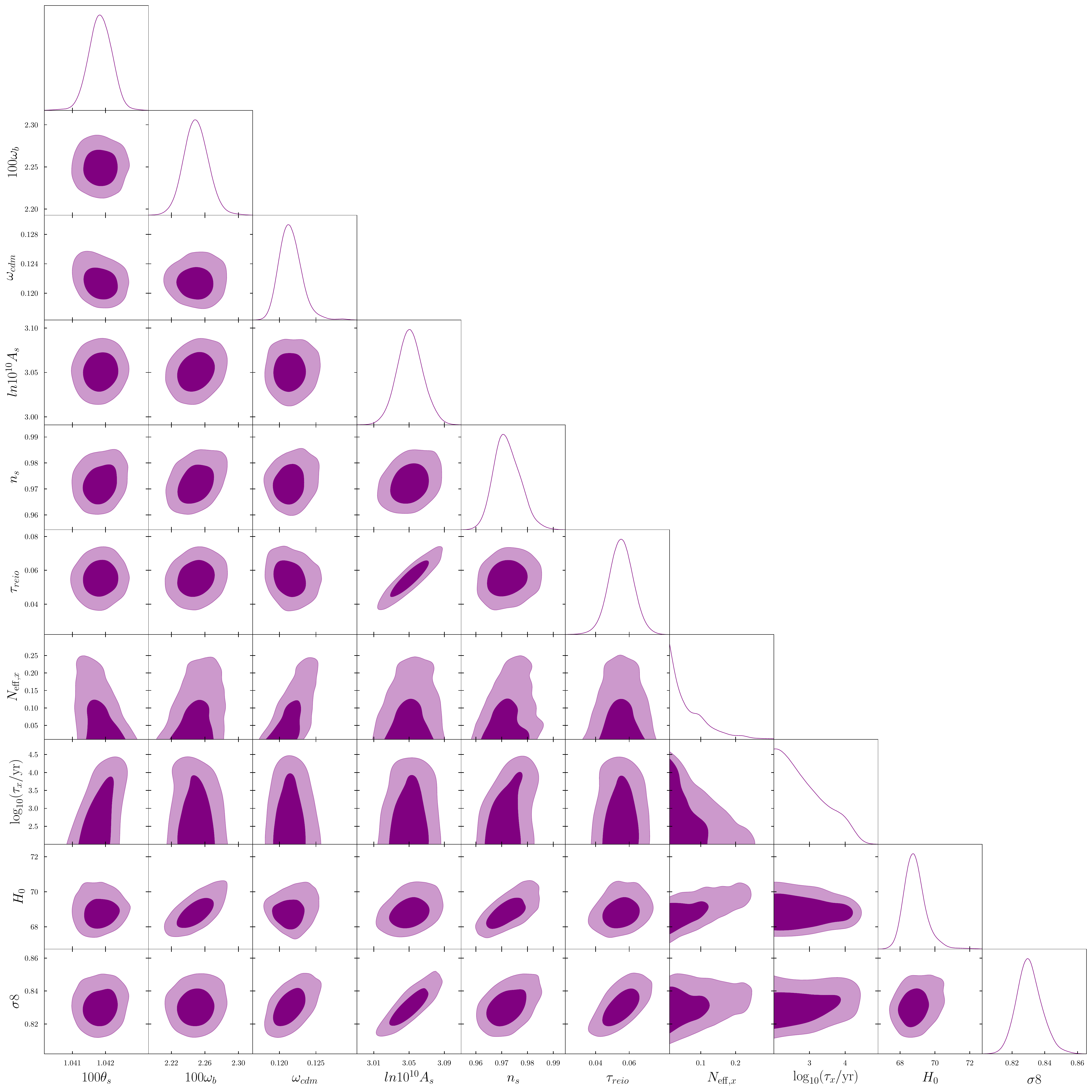}
    \caption{Posterior distributions for the DWDM model with $m_x =10$ eV, for the Planck + BAO data set. Inner darker (outer lighter) regions correspond to 
    $1\sigma$ ($2\sigma$) confidence regions.}
    \label{fig:bao10_tri}
\end{figure}

\begin{figure}
    \centering
    \includegraphics[width=\textwidth]{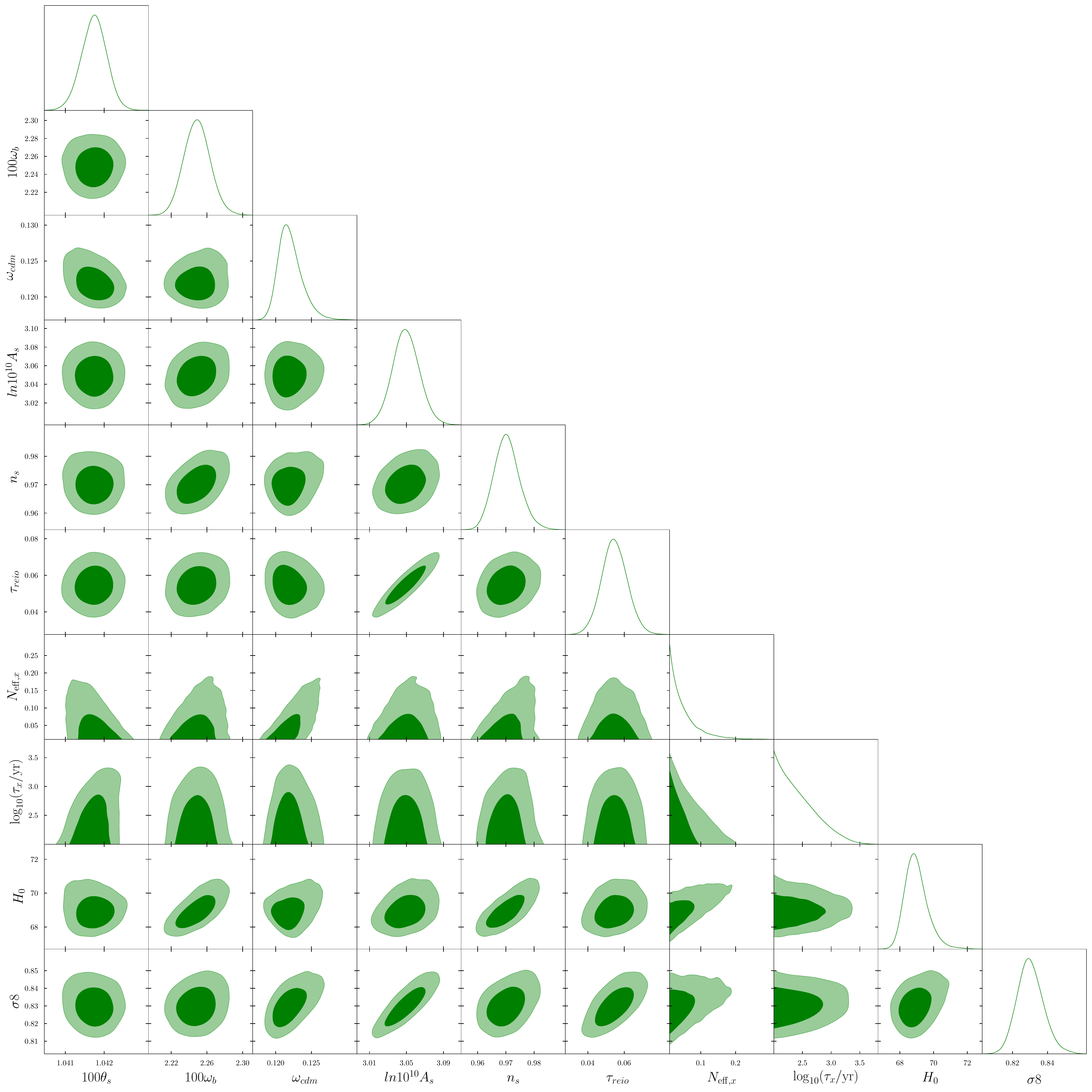}
    \caption{Posterior distributions for the DWDM model with $m_x =40$ eV, for the Planck + BAO data set. Inner darker (outer lighter) regions correspond to 
    $1\sigma$ ($2\sigma$) confidence regions.}
    \label{fig:bao40_tri}
\end{figure}

Lastly, in this appendix we present the full set of 2D posterior distributions for the DWDM model with $m_x = 1$ eV (Fig.~\ref{fig:bao1_tri}), 
$10$ eV (Fig.~\ref{fig:bao10_tri}), and $40$ eV (Fig.~\ref{fig:bao40_tri}). In each of these figures, we show results for the Planck + BAO data set, as described in Sec.~\ref{sec:results}.

\bibliographystyle{JHEP}
\bibliography{biblio}
\end{document}